\documentclass[11pt]{article}
\pdfoutput=1

\usepackage{jheppub}
\usepackage{amssymb,amsfonts}
\usepackage{enumerate}
\usepackage{mathrsfs}

\newcommand{\be}{\begin{equation}}
\newcommand{\ee}{\end{equation}}
\newcommand{\bea}{\begin{eqnarray}}
\newcommand{\eea}{\end{eqnarray}}

\newtheorem{Identity}{Identity}

\newcommand{\bra}[1]{\langle #1|}
\newcommand{\ket}[1]{|#1\rangle}
\newcommand{\corr}[1]{\langle{#1}\rangle}

\newcommand{\R}{\mathbb{R}}

\newcommand{\qD}{\mathcal{D}}
\newcommand{\qO}{\mathcal{O}}
\newcommand{\cZ}{\mathcal{Z}}

\newcommand{\CD}{{\cal D}}

\newcommand{\CG}{{\cal G}}

\newcommand{\CI}{{\cal I}}
\newcommand{\CJ}{{\cal J}}

\newcommand{\CN}{{\cal N}}

\newcommand{\CO}{{\cal O}}
\newcommand{\CR}{{\cal R}}

\newcommand{\CV}{{\cal V}}

\newcommand{\CZ}{{\cal Z}}

\newcommand{\bk}{{\bf k}}

\newcommand{\bx}{{\bf x}}

\newcommand{\lr}{\left (}
\newcommand{\rr}{\right )}
\newcommand{\ls}{\left [}
\newcommand{\rs}{\right ]}
\newcommand{\lc}{\left \{}
\newcommand{\rc}{\right \}}

\newcommand{\zN}{\overline{N}}
\newcommand{\ze}{\overline{\varepsilon}}
\newcommand{\zg}{\overline{g}}
\newcommand{\zK}{\overline{K}}
\newcommand{\zR}{\overline{R}}
\newcommand{\zB}{\overline{\square}}
\newcommand{\zD}{\overline{\nabla}}

\newcommand{\qac}{\mathfrak{u}}
\newcommand{\qao}{\mathfrak{u}_1}
\newcommand{\qat}{\mathfrak{u}_2}

\newcommand{\qth}{\mathfrak{v}}
\newcommand{\qM}{S^{(2)}}

\newcommand\qs{s}
\newcommand\qt\tau

\newcommand{\req}[1]    {(\ref{#1})}

\newcommand{\p}{\partial}

\renewcommand{\bar}[1]{\overline{#1}}
\renewcommand{\tilde}[1]{\widetilde{#1}}
\newcommand{\tr}{\text{tr}}

\makeatletter
\renewcommand{\@seccntformat}[1]{\csname the#1\endcsname.\,\,}
\makeatother

\let \savenumberline \numberline
\def \numberline#1{\savenumberline{#1.}}

\makeatletter
\def\@fpheader{\relax}
\makeatother

\title{\ \vspace{1.6cm} \\
Quantization of Ho\v{r}ava Gravity in 2+1 Dimensions}
\author{Tom Griffin${}^a$, Kevin T. Grosvenor${}^b$, 
Charles M. Melby-Thompson${}^{c,d}$, and Ziqi Yan${}^{e,f}$}
\emailAdd{t.griffin@imperial.ac.uk} 
\emailAdd{kevin.grosvenor@nbi.ku.dk}
\emailAdd{charlesmelby@gmail.com}
\emailAdd{zyan@berkeley.edu}
\affiliation{${}^a$Blackett Laboratory, Department of Physics\\
Imperial College, London, SW7 2AZ, UK\medskip\\
${}^b$Niels Bohr Institute, University of Copenhagen\\
Blegdamsvej 17, DK-2100 Copenhagen \O, Denmark\medskip\\
${}^c$Department of Physics, Fudan University\\
220 Handan Road, 200433 Shanghai, China\medskip\\
${}^d$Kavli Institute for the Physics and Mathematics of the Universe (WPI)\\
The University of Tokyo Institutes for Advanced Study (UTIAS)\\
The University of Tokyo, Kashiwanoha, Kashiwa, 277-8583, Japan\medskip\\
${}^e$Berkeley Center for Theoretical Physics and Department of Physics\\
University of California, Berkeley, CA 94720-7300, USA\medskip\\
${}^f$Theoretical Physics Group, Lawrence Berkeley National Laboratory\\
Berkeley, CA 94720-8162, USA}
\abstract{
We study quantum corrections to projectable Ho\v{r}ava gravity with $z=2$ scaling in $2+1$ dimensions.
Using the background field method, we utilize a non-singular gauge to compute the anomalous dimension of the cosmological constant at one loop, in a normalization adapted to the spatial curvature term.
}

\begin{document}
\maketitle

\section{Introduction}

Considerable effort has been devoted to the study of Ho\v{r}ava gravity since it was introduced in \cite{mqc,qglp}. 
Being renormalizable by na\"ive power counting, Ho\v{r}ava's theory constitutes a candidate for an ultraviolet-complete theory of quantum gravity.
In spite of some work \cite{Saueressig, semiHL, Reffert, reHL}, nonetheless, as yet there have been no fully satisfactory quantum computations;
in fact, perturbative renormalizability of one version --- the ``projectable'' model --- was established only recently~\cite{Blas}.

The purpose of the present paper is to take a step forward in understanding quantum corrections to Ho\v{r}ava gravity by making a careful computation of a one-loop quantity working in non-singular gauges.
(What we mean by this is explained in Sections~\ref{sec:flat space} and~\ref{sec:BRSTquantization}.)
More specifically, the model we consider is $z=2$ projectable Ho\v{r}ava gravity in $2+1$ dimensions, and the quantity we compute is the anomalous dimension of the cosmological constant.%
\footnote{In anisotropic models, the effective coefficients of the temporal and spatial kinetic terms can scale differently --- \emph{i.e.,} the dispersion relation runs with scale. 
This running can be captured by fixing the form in which either the energy or the spatial momentum appears in the dispersion relation. 
We compute the anomalous dimension of $\Lambda$ with respect to a normalization condition that fixes the form of the spatial momentum contribution.}

Ho\v{r}ava gravity is constructed so that, at high energies, the classical action has anisotropic scale invariance with the dynamical critical exponent $z$:
\be
	t \rightarrow b^z \, t,
	\qquad \mathbf{x} \rightarrow b \, \mathbf{x}.	
\ee
As our interest is in $z=2$, we take the engineering dimensions of the time and space coordinates to be
\be
	[t] = -1, 
	\qquad [\mathbf{x}] = - \frac{1}{2}.
\ee 
In this convention, energy is of dimension one.  
The $z=2$ theory is renormalizable in $2+1$ dimensions~\cite{Blas}.

The spacetime manifold is equipped with a foliation by leaves of codimension one, corresponding to the surfaces of constant time.
Its geometry is naturally parametrized using the ADM variables -- a spatial scalar $N$ (the lapse), a spatial vector $N_i$ (the shift), and a spatial metric $g_{ij}$. The classical scaling dimensions of the fields are
\be
	[N] = 0, 
		\qquad
	[N_i] = \frac{1}{2},
		\qquad
	[g_{ij}] = 0.
\ee
The gauge symmetries are the diffeomorphisms that preserve the foliation.
We parametrize the infinitesimal transformations by $(Z,X^i)$,
\be
	\delta t = Z (t), 
		\qquad
	\delta x^i = X^i (t, \mathbf{x}) \,,
\ee
that act on the fields by
\begin{subequations} \label{eq:gauge transformations}
\begin{align}
    \delta N & = \p_t(Z\,N) + X^k\nabla_k N, \label{eq:delta N} \\
    \delta N^i &= \p_t(Z\,N^i) + (\p_t-N^k\nabla_k)X^i + X^k\nabla_kN^i, \label{eq:delta N^i} \\
    \delta g_{ij} &= Z\dot g_{ij} + \nabla_i X_j + \nabla_j X_i. \label{eq:delta g_ij}
\end{align}
\end{subequations}

A proper understanding of Ho\v{r}ava gravity requires a careful treatment of its gauge fixing.
To this end, it is useful to begin with the simplest model possible. 
It is tempting to begin with the conformal case in $2+1$ dimensions, because it has no local propagating degrees of freedom.
Unfortunately, not only does it require the ``non-projectable'' version of the theory, which has second class constraints and their attendant difficulties, but also it raises the thorny issue of gauge anomalies for the Weyl symmetry.

A more modest starting point is ``projectable'' Ho\v{r}ava gravity in $2+1$ dimensions. 
Projectability is the condition that $N=N(t)$ be a function of time but not of space, so that it is constant on each spatial slice. 
We assume this condition for the remainder of the paper.
The $2+1$ dimensional projectable case is more than just a toy model for understanding the qualitative behavior of the more realistic $3+1$ dimensional non-projectable theory. Mapping out the renormalization group (RG) structure of the projectable theory is important to further understand the phases of gravity, both in the context of Ho\v{r}ava gravity and the Causal Dynamical Triangulation approach to quantum gravity \cite{CDT,gcglp}.

The action is written in terms of quantities invariant under those diffeomorphisms preserving the foliation of spacetime, namely scalars built from the intrinsic and extrinsic curvatures of the leaves of the foliation and their covariant derivatives. The intrinsic curvature of a two-dimensional leaf is completely determined by its spatial Ricci scalar $R$. The extrinsic curvature is captured by the tensor
\be
    K_{ij} = \frac{1}{2N} ( \dot{g}_{ij} - \nabla_i N_j - \nabla_j N_i ),
\ee
where $\nabla_i$ is the covariant derivative with respect to $g_{ij}$. The most general $z=2$ action invariant under~\req{eq:gauge transformations} is
\be \label{eq:action}
	S = \frac{1}{\kappa^2} \int dt \, d^2 \mathbf{x} \, N \sqrt{g} \, \Big \{ K_{ij} K^{ij} - \lambda K^2 - \gamma R^2 + \rho R - 2 \Lambda \Big \},
\ee
where $K = g^{ij} K_{ij}$. 
Since $\int \! d^2 \mathbf{x} \sqrt{g} R$ is a topological invariant in two dimensions, $\rho$ does not appear in the local equations of motion, but only in the global Hamiltonian constraint arising from time reparametrization symmetry.
As a result $\rho$ cannot contribute to the perturbative beta function, and so we drop this term in what follows.%
\footnote{On the other hand, it may very well contribute to the full non-perturbative beta function through instanton corrections. Also note that, while it cannot contribute to the perturbative beta function, in principle $\rho$ itself may have a non-zero perturbative beta function that depends only on the other couplings in the theory. For dimensional reasons, however, its beta function vanishes at one loop. (See Section \ref{sec:renorm}.)}

In general dimension, projectable Ho\v{r}ava gravity has a transverse traceless tensor mode and a scalar mode.
Requiring the tensor polarizations to have a good dispersion relation around flat space then implies that $\gamma>0$. 
Requiring the dispersion of the scalar also to be healthy imposes the constraint
\be \label{eq:ulam}
	\lambda < \frac{1}{2} 
		\quad \text{or} \quad
	\lambda > 1.
\ee
In $2+1$ dimensions, however, there are no tensor modes.
We then have the option of  
setting $\gamma$ to be negative when $\frac{1}{2} < \lambda < 1$. The propagating spectrum of the theory is then healthy, at least classically. We do not worry about this explicitly in what follows, although our final result makes sense in this parameter region.

In this paper, we will compute contributions to the effective action using the background field method.
In this method, fields are split into a sum of two terms: a classical background value, and quantum fluctuations of typical size $\hbar^{1/2}$.
For the action~\req{eq:action}, the role of $\hbar$ is played by $\kappa^2$.
This leads us to expand
\be \label{eq:background_expansion}
	N = \zN + \kappa\, n, \qquad N^i = \zN^i + \kappa\, n^i, \qquad g_{ij} = \zg_{ij} + \kappa\, h_{ij},
\ee
where $\zN$, $\zN^i$ and $\zg_{ij}$ are background fields and $n$, $n^i$ and $h_{ij}$ are fluctuations around the given background. 
Gauge transformations can also be expanded in powers of $\kappa$,
\be
    Z = \bar Z + \kappa\,\zeta,
    \qquad
    X^i = \bar X^i + \kappa\,\xi^i,
\ee
with $(\bar Z,\bar X^i)$ the background diffeomorphisms, and $(\zeta,\xi^i)$ the physical gauge symmetries of the quantum fluctuations.
Due to the projectability condition, we can use $(\bar Z,\bar X^i)$ to set
\be
	\zN = 1, \qquad \zN^i = 0.
\ee
In this gauge, the action of $\zeta$ and $\xi^i$ (to linear order in $\kappa$) is
\begin{subequations}\label{eq:gauge}
\begin{align}
    \delta n &= \dot\zeta + O(\kappa), \\
    \delta n^i &= \dot \xi^i + O(\kappa), \\
    \delta h_{ij} &= \zD_{\!i} \xi_j + \zD_{\!j} \xi_i + O(\kappa) \,.
\end{align}
\end{subequations}
Here, $\zD_i$ denotes the Christoffel connection for $\zg_{ij}$.
We can use $\zeta$ to set $n \equiv 0$; since $n$ is independent of space and so has only one degree of freedom per spatial slice, it does not contribute divergences.
For our purposes, therefore, we can ignore the contribution from $n$ to the path integral.

\bigskip
In the following, we will work only on backgrounds that are time-independent.
We express the partition function in terms of functional determinants by integrating out the quantum fluctuations $n_i$ and $h_{ij}$, and the gauge-fixing ghost modes. The one-loop effective action is then evaluated using heat kernel techniques. This will allow us to compute some (but not all) of the one-loop beta functions in the theory. To fully understand the RG properties of the theory at weak coupling (and in particular, determine whether the theory is asymptotically free), it is necessary to evaluate the heat kernel on background geometries with a time-dependent metric. We leave this to future work.

Previous work on the one-loop effective action in gravity with anisotropic scaling~\cite{Benedetti} overlooked crucial contributions from the gauge-fixing sector of the theory, a problem exacerbated by dropping from the partition function altogether singular determinants that did not cancel out in their analysis.
We show that such confusion can be avoided by an appropriate choice of gauge. 
The gauge-fixing methods we developed have, in the meantime, appeared in a more general form in the work of~\cite{Blas}, which applied them to show the renormalizablility of projectable Ho\v{r}ava gravity.
We take advantage of their more general gauge in Section \ref{sec:flat space} for reasons of clarity, although the bulk of our computation uses our more restrictive original gauge.

Section~\ref{sec:anisotropic gauge fixing} develops the gauge-fixing method and field parametrizations we use in the remainder of the paper in the simpler context of linearized theories.
Before embarking on the gravitational calculation, we begin in Section~\ref{sec:U1} with a warm-up -- free $U(1)$ gauge theory in $D+1$ dimensions with $z=2$ scaling at short distances. 
One natural choice in this context, used in \cite{qcYM}, is temporal gauge. 
Here, we utilize a gauge choice that manifestly respects the $z=2$ scaling symmetry. 
Generalizing this gauge-fixing procedure to the gravitational case will lead us in Section~\ref{sec:flat space} to the same sort of gauge-fixing condition used by~\cite{Blas} in proving perturbative renormalizability of projectable Ho\v{r}ava gravity. 
Section~\ref{sec:flat space loop} uses these results to compute the dependence of the one-loop effective action on the cosmological constant, which illustrates how the effective action can depend on gauge, and how to extract the correct gauge-invariant effective action.

Section~\ref{sec:curved background} turns to computations in curved space using the background field method.
There, we compute the partition function on static on-shell curved backgrounds ($\zR = \mbox{const}$, $\p_t\zg_{ij} = 0$) supported by non-vanishing $\Lambda$.
Working with an on-shell background enables us to systematically disentangle the physical and unphysical modes and observe explicitly the cancellation of the unphysical modes among themselves. We give an explicit expression for the physical dispersion relation, which generalizes the flat space result.
We normalize the gravitational field such that $\gamma / \kappa^4$ is constant at all energy scales.
With respect to this choice of normalization condition, we are able to determine the anomalous dimension of $\Lambda$.
Extracting the beta functions for $\gamma$, $\lambda$ and $\kappa$ requires working on backgrounds that depend on time, which we leave to future research.

\section{Gauge Fixing in Theories with Anisotropic Scaling}
\label{sec:anisotropic gauge fixing}

In gauge theories exhibiting 
an anisotropic scaling symmetry of the form $(t, \mathbf{x} ) \mapsto (b\,t, b^{1/z} \mathbf{x} )$,
it is desirable to choose a gauge-fixing condition that respects this symmetry. 
This is especially true in models at their critical dimension, 
for which standard gauges -- in particular, Lorenz gauge -- may not be renormalizable.

In some simple cases (\textit{e.g.,} free Maxwell theory), there is no problem with singular gauges, such as the temporal or Coulomb gauges, which are in fact invariant under the scaling symmetry for any value of $z$. 
When the theory is coupled to gravity, however, such gauges can become problematic.
For example, in temporal gauge the Faddeev-Popov determinant is $\det ( \p_t )$.
While in the flat case this determinant can be dropped, in the gravitational case it couples non-trivially and should not be ignored.
However, such operators have no dependence on large spatial momenta, leading to uncontrolled ultraviolet divergences.
Moreover, this problem persists in both dimensional regularization and heat-kernel based methods.
Such gauges therefore give rise to ambiguities, which need to be resolved in a manner consistent with BRST symmetry.
From a more pedestrian perspective, our strategy ensures that the gauge-fixing Lagrangian, which is quadratic in the gauge-fixing condition, is of the same order in derivatives as the original Lagrangian. 
Thus, the two can be combined more seamlessly.

In this section, our goal is to introduce%
\footnote{The gauges we use in this paper also appeared in the work of~\cite{Blas}, where they were used to demonstrate the perturbative renormalizability of projectable Ho\v{r}ava gravity. We originally arrived at them as a way to remove singular behavior in the background field formalism while preserving anisotropic Weyl invariance.}
such gauges in linearized $z=2$ gravity.
We first illustrate the process in free anisotropic $U(1)$ gauge theory.
This serves as a warm-up to the second case of $z=2$ projectable Ho\v{r}ava gravity in $2+1$ dimensions linearized around flat space. 
We apply these results to make a simple quantum computation.
Section \ref{sec:curved background} will be concerned with the generalization to static backgrounds in the background field method.

\subsection{\texorpdfstring{$U(1)$}{U(1)} gauge theory}
\label{sec:U1}

We begin with free $U(1)$ gauge theory in $D+1$ dimensions exhibiting $z=2$ scaling in the ultraviolet (UV) and $z=1$ in the infrared (IR). 
The gauge field is a $U(1)$ connection on Aristotelian spacetime \cite{Higgs}. 
The time and space components, $A_0$ and $A_i$ ($i = 1, \cdots, D$), have gauge transformations,
\be
	\delta A_0 = \dot{\zeta},
		\qquad
	\delta A_i = \partial_i \zeta,
\ee 
with $\zeta(t,\bx)$ an arbitrary scalar function.
The invariant field strengths are
\be
	E_i = \dot{A}_i - \partial_i A_0,
		\qquad
	F_{ij} = \partial_i A_j - \partial_j A_i.
\ee
At the ultraviolet $z=2$ Gaussian fixed point,
the engineering dimensions of the gauge fields are
\be
	[A_0] = \frac{D}{4}, 
		\qquad
	[A_i] = \frac{D}{4}-\frac{1}{2}.
\ee
The basic most generic 
action with this scaling in the UV that is invariant under both the 
spacetime and 
gauge symmetries (including parity and time-reversal) is
\be
	S = \int dt \, d^D \mathbf{x} \Big \{ \frac{1}{2} E_i E_i - \frac{1}{4}\partial_k F_{ij} \partial_k F_{ij} - \frac{1}{4}v^2 F_{ij} F_{ij} \Big \},
\ee
where $v$ is the ``speed of light'' in the infrared. 

In components, the action becomes
\begin{align} \label{eq:U1S}
	S &= \frac{1}{2} \int dt \, d^D \mathbf{x} \Big \{ \partial_i A_0 \partial_i A_0 + \dot{A}_i \dot{A}_i - 2 \dot{A}_i \partial_i A_0 - A_i \lr \partial^2 - v^2 \rr \lr \delta_{ij} \partial^2 - \partial_i \partial_j \rr A_j \Big \} \notag \\
	&= \frac{1}{2} \int dt \, d^D \mathbf{x} 
\begin{pmatrix} A_0 & A_i \end{pmatrix}
    \qM
\begin{pmatrix}
    A_0 \\
    A_j
\end{pmatrix},
\end{align}
where
\be \label{eq:U2}
    \qM = 
\begin{pmatrix}
    - \partial^2 \quad & \partial_j \partial_t \\
    \partial_t \partial_i \quad & \qO \delta_{ij} + ( \partial^2 - v^2 ) \partial_i \partial_j 
\end{pmatrix},
\ee 
and $\qO$ is the generalized d'Alebertian operator,
\be
    \qO = - \partial_{t}^{2} - \partial^4 + v^2 \partial^2.
\ee

A natural $z=2$ generalization of the Lorenz gauge is given by the gauge-fixing functional\footnote{Quantization of anisotropic gauge theory using a gauge-fixing functional of this form was first studied in \cite{Anselmi}.}
\be \label{eq:gfcond}
    f[A] = \dot{A}_0 - ( - \partial^2 + v^2 ) \partial_i A_i.
\ee
To quantize the theory with this gauge-fixing, we should further introduce a pair of fermionic ghosts ($b,c$), a bosonic auxiliary field $\Phi$, and the fermionic BRST differential $s$ acting on the fields as
\begin{align}
    sA_0 &= \dot{c}, &%
    sA_i &= \partial_i c, &%
    s b &= \Phi, &%
    s \Phi &= sc = 0.
\end{align}
A generalized
$\CR_\xi$ gauge-fixing action based on~\eqref{eq:gfcond} can now be obtained from a gauge-fixing fermion of the form
\be
	\Psi = \int dt \, d^D \mathbf{x} \, b \, \Bigl\{ \frac{1}{2} \qD \Phi - f[A] \Bigr\}
	\,.
\ee
Note that, unlike standard $\CR_\xi$ gauge, if we wish to avoid introducing dimensionful parameters then $\qD$ must be a differential operator of dimension one. 
The BRST-exact action is
\be \label{eq:BRSTexactaction}
    s \Psi = \int dt \, d^D \mathbf{x} \, \Big \{ \frac{1}{2} \Phi \qD \Phi - \Phi f[A] + b \qO c \Big \}.
\ee
The resulting BRST-invariant gauge-fixed action is
\be
    S_{\text{BRST}} = S + s \Psi
    \,,
\ee
giving the quantum partition function
\be \label{eq:U1Z}
    \mathcal{Z} = \int \mathscr{D} \{ A_0, A_i, b, c, \Phi \} \, e^{iS_{\text{BRST}}}
    \,.
\ee
As in the case of the standard $\CR_\xi$-gauge procedure, the partition function is independent of $\qD$. We demonstrate this explicitly in Appendix \ref{app:U1}.

Setting $\qD = - \partial^2 + v^2$ is a particularly nice choice, as it eliminates the cross-terms between $A_0$ and $A_i$, after integrating out $\Phi$. 
Redefining the $A_0$ field via $A_0 \rightarrow \sqrt{\qD} \, A_0$ results in a Jacobian $\CJ_{A_0}=( \det \qD )^{1/2}$, which cancels the factor of $(\det \qD )^{-1/2}$ produced by the integral over $\Phi$. 
The action then becomes
\be \label{eq:SBRSTA0}
    S_{\text{BRST}} [A_0, A_i, b ,c ] = \int dt \, d^D \mathbf{x} \Big \{ \! - \frac{1}{2} A_0 \qO A_0 + \frac{1}{2} A_i \qO A_i + b \qO c \Big \}.
\ee
The overall sign in front of the piece quadratic in $A_0$ in \eqref{eq:SBRSTA0} is negative, so we must Wick rotate $A_0$ when we rotate $t$.
The partition function evaluates to
\be
    \mathcal{Z} = \lr \det \qO \rr^{- \frac{D-1}{2}}.
\ee
This represents $D-1$ physical propagating modes with dispersion relation
\be \label{eq:U1dispersion}
    \omega^2 = k^4 + v^2 k^2.
\ee

Before we move on to Ho\v{r}ava gravity, we make the following comment. As in the Lorentz-invariant theory, one can diagonalize the kinetic operator of \eqref{eq:U1S} explicitly in field space, without gauge-fixing.
There is one pure gauge mode on which the operator vanishes completely.
There is also one unphysical mode which gets a wrong-sign dispersion relation. The rest of the modes should then reproduce the correct dispersion relation \eqref{eq:U1dispersion}. We perform this exercise for illustrative purposes in Appendix \ref{app:direct_eigenvalues}.\footnote{K.T.G. would like to thank Laure Berthier for this point.}

\subsection{Ho\texorpdfstring{\v{r}}{r}ava gravity around flat space}
\label{sec:flat space}

We now turn to the linearization of $z=2$ projectable Ho\v{r}ava gravity in $(2+1)$ dimensions around flat space, with $\overline{g}_{ij}$ in \eqref{eq:background_expansion} set to $\delta_{ij}$.
The flat background is on-shell if the cosmological constant $\Lambda$ is set to zero. 
However, since we are also interested in the $\Lambda$ dependence of the off-shell effective action, we will allow for a nonzero $\Lambda$.

The action is that of \eqref{eq:action}, with $\rho = 0$. 
The quadratic part%
\footnote{Since we are interested in the effective action, we drop the linear part, which is non-vanishing when $\Lambda\ne 0$.}
of the action is
\begin{align} \label{eq:HGflatS}
	S_\text{quad} &= \int dt \, d^2 \mathbf{x} \Big \{ \frac{1}{4} \lr \dot{h}_{ij} \dot{h}_{ij} - \lambda \dot{h}^2 \rr - \dot{h}_{ij} \partial_i n_j + \lambda \, \dot{h} \partial_i n_i + \frac{1}{2} \partial_i n_j \partial_i n_j - \lr \lambda - \frac{1}{2} \rr \lr \partial_i n_i \rr^2 \notag \\
	&\hspace{7.15cm} - \gamma (\partial_i \partial_j h_{ij} - \partial^2 h)^2 
	+ \frac{\Lambda}{4} ( 2 h_{ij} h_{ij} - h^2) \Big \} \notag \\
	&= \frac{1}{2} \int dt \, d^2 \mathbf{x} \,
\begin{pmatrix}
    n_i & h_{ik}
\end{pmatrix}
    \qM
\begin{pmatrix}
    n_j \\
    h_{j \ell}
\end{pmatrix},
\end{align}
where $h \equiv h_{ii}$, and the matrix $\qM$ is the second functional derivative of the action.
Explicitly,
\be \label{eq:HGM}
    \qM = 
\begin{pmatrix}
    \qM_{nn} \,\, & \qM_{nh} \\[4pt]
    \qM_{hn} \,\, & \qM_{hh}
\end{pmatrix},
\ee 
with
\begin{subequations} \label{eq:HGMs}
\begin{align}
    \qM_{nn} &= - \delta_{ij} \partial^2 + ( 2 \lambda - 1 ) \partial_i \partial_j, \label{eq:Mnn} \\
    \qM_{nh} &= \ls \qM_{hn} \rs^{\dagger} \! = \frac{1}{2} \bigl( \delta_{ij} \partial_{\ell} + \delta_{i \ell} \partial_j - 2 \lambda \delta_{j \ell} \partial_i \bigr) \partial_t, \label{eq:Mnh} \\
    \qM_{hh} &= - \frac{1}{4} \bigl( \delta_{ij} \delta_{k \ell} + \delta_{i \ell} \delta_{jk} - 2 \lambda \delta_{ik} \delta_{j \ell} \bigr) \partial_{t}^{2} + \frac{\Lambda}{2} ( \delta_{ij} \delta_{k \ell} + \delta_{i \ell} \delta_{jk} - \delta_{ik} \delta_{j \ell} ) \notag \\
    &\quad - 2 \gamma \ls \delta_{ik} \delta_{j \ell} \partial^4 - ( \delta_{ik} \partial_j \partial_{\ell} + \delta_{j \ell} \partial_i \partial_k ) \partial^2 + \partial_i \partial_j \partial_k \partial_{\ell} \rs.
\end{align}
\end{subequations}
Intuitively, one can think of this theory as ``adding a spatial index" to the $U(1)$ gauge theory of the previous section: $n_i$ is analogous to $A_0$, and $h_{ij}$ to $A_i$. Likewise, the gauge-fixing functional $f_i$, ghost fields $b_i$ and $c_i$, and bosonic auxiliary field $\Phi_i$ all carry a spatial index. 
The BRST differential $s$ acts as
\begin{align}
    sn_i &= \dot{c}_i, &%
    sh_{ij} &= \partial_i c_j + \partial_j c_i, &%
    s b_i &= \Phi_i, &%
    s \Phi_i &= sc_i = 0.
\end{align}	
In analogy with the $U(1)$ theory, we choose the gauge-fixing fermion 
\be
	\Psi = \int dt \, d^2 \mathbf{x} \, b_i \Big \{ \frac{1}{2} \qD_{ij} \Phi_j - f_i
	\Big \},
\ee
where $\qD_{ij}$ is some spatial differential operator of dimension one, and $f_i$ is a gauge-fixing functional.
As pointed out in \cite{Blas}, the most general such operator is
\be \label{eq:Dij}
    \qD_{ij} = - \qao \delta_{ij} \partial^2 - \qat\, \partial_i \partial_j
    \,,
\ee 
where $\qao$ and $\qat$ are constants. 

The analog of the gauge-fixing condition~\eqref{eq:gfcond} reads $f_i = \dot{n}_i - \mathcal{D}_{ijk} h_{jk}$, where $\mathcal{D}_{ijk}$ is a spatial differential operator of energy dimension $\frac{3}{2}$ (e.g., containing three spatial derivatives). 
Forcing the cross-terms between $n_i$ and $h_{ij}$ to vanish upon integrating out $\Phi_i$ uniquely determines 
$\mathcal{D}_{ijk}$ to be $\mathcal{D}_{ijk} = \mathcal{D}_{ij} \partial_k - \lambda \mathcal \delta_{jk} {\mathcal{D}}_{i \ell} \partial_{\ell}$:
\begin{align} \label{eq:flatfic}
    f_i &= \dot{n}_i - \qD_{ij} ( \partial_k h_{jk} - \lambda \partial_j h ) \notag \\
    &= \dot{n}_i + \qao \, \partial^2 \partial_j h_{ij} + \qat \, \partial_i \partial_j \partial_k h_{jk} - \lambda \qac \, \partial^2 \partial_i h,
\end{align}
where $\qac = \qao + \qat$.
As before, the final result is independent of the particular choice of $\mathcal{D}_{ijk}$.

The analog of the BRST-exact action \eqref{eq:BRSTexactaction} is
\be
    S' = s \Psi = \int dt \, d^2 \mathbf{x} \, \Big \{ \frac{1}{2} \Phi_i \qD_{ij} \Phi_j - \Phi_i f_i [h,n] + b_i \qO_{ij} c_j \Big \},
\ee
where
\begin{align}
    \qO_{ij} &= - \delta_{ij} \partial_{t}^{2} + \qD_{ik} \bigl[ \delta_{jk} \partial^2 - ( 2 \lambda - 1 ) \partial_j \partial_k \bigr] \notag \\
    &= \delta_{ij} \bigl( - \partial_{t}^{2} - \qao \partial^4 \bigr) + 2 \bigl[ \bigl( \lambda - \tfrac{1}{2} \bigr) \qao + ( \lambda - 1 ) \qat \bigr] \partial^2 \partial_i \partial_j.
\end{align}
We can immediately read off the ghost partition function,
\be
\label{eq:Zghflat}
    \mathcal{Z}_\text{ghost} = \det \mathcal{O}_{ij} = (\det \mathcal{O}_g) (\det \tilde{O}_g),
\ee
where
\begin{align} 
    \mathcal{O}_g &= - \partial_{t}^{2} - 2 \qac ( 1 - \lambda ) \partial^4, &%
    \tilde{\mathcal{O}}_g &= - \partial_{t}^{2} - \qao \partial^4.
\end{align}
The rest of the action, after integrating out $\Phi_i$, called the ``effective" part, reads
\be \label{eq:Seffn0}
    S_{\text{eff}} [ n_i , h_{ij} ] = \frac{1}{2} \int dt \, d^2 \mathbf{x} \, \Big \{ \! - n_i \qM_{ij} n_j + h_{ij} \qM_{ijk \ell} h_{k \ell} \Bigr \},
\ee
where
\begin{subequations}
\begin{align}
    \qM_{ij} &= \qD^{-1}_{ik} \qO_{kj}, \\
    \qM_{ijk \ell} &= \frac{1}{4} ( \delta_{ik} \delta_{j \ell} + \delta_{i \ell} \delta_{jk} ) ( - \partial_{t}^{2} + 2 \Lambda ) - \frac{1}{4} \delta_{ij} \delta_{k \ell} ( -2 \lambda \partial_{t}^{2} - 4 \lambda^2 \qD_{mn} \partial_m \partial_n + 8 \gamma \partial^4 + 2 \Lambda ) \notag \\
    &\quad -2 \gamma \partial_i \partial_j \partial_k \partial_{\ell} + 2 \gamma ( \delta_{ij} \partial_k \partial_{\ell} + \delta_{k \ell} \partial_i \partial_j ) \partial^2 \notag \\
    &\quad + \frac{1}{4} ( \qD_{ik} \partial_j \partial_{\ell} + \qD_{i \ell} \partial_j \partial_k + \qD_{jk} \partial_i \partial_{\ell} + \qD_{j \ell} \partial_i \partial_k ) \notag \\
    &\quad - \frac{\lambda}{2} \bigl[ \delta_{ij} ( \qD_{km} \partial_{\ell} + \qD_{\ell m} \partial_k ) + \delta_{k \ell} ( \qD_{im} \partial_j + \qD_{jm} \partial_i ) \bigr] \partial_m. 
\end{align}
\end{subequations}
Note that various field components need to be Wick-rotated
as well as the time when performing the path integral.
The contribution of $n_i$ to the partition function is
\be \label{eq:Zn}
     \mathcal{Z}_n = \bigl( \det \qM_{ij} \bigr )^{-1/2} = \lr \det \CD_{ik} \rr^{1/2} \lr \det \qO_{ij} \rr^{-1/2}.
\ee

Next, we compute the contribution to the partition function from $h_{ij}$. We first decompose $h_{ij}$ as
\be \label{eq:htoH}
    h_{ij} = H_{ij} + \frac{1}{2} h \delta_{ij},
\ee
where $H_{ii} = 0$. We decompose $H_{ij}$ further as
\be \label{eq:Hij1}
    H_{ij} = H^\perp_{ij} + \partial_i \eta_j + \partial_j \eta_i + \biggl( \partial_i \partial_j - \frac{1}{2} \delta_{ij} \partial^2 \biggr) \sigma,
\ee
with constraints
\begin{align}
    H^\perp_{ii} &= 0, &%
    \partial_j H^\perp_{ij} &= 0, &%
    \partial_i \eta_i &= 0.
\end{align}
In two spatial dimensions, the transverse traceless component $H^\perp_{ij}$ has no local degrees of freedom, and in flat space is forced by boundary conditions to vanish.
Furthermore, in two dimensions, one can parametrize $\eta_i$ as
\be 
    \eta_i = \epsilon_{ij} \partial_j \eta.
\ee
Thus, $H_{ij}$ is finally parametrized as
\be \label{eq:Hij2}
    H_{ij} = \lr \epsilon_{ik} \partial_j + \epsilon_{jk} \partial_i \rr \partial_k \eta + \biggl( \partial_i \partial_j - \frac{1}{2} \delta_{ij} \partial^2 \biggr) \sigma.
\ee
We require the Jacobian for this change of variables. The Jacobian for \eqref{eq:htoH} is a constant. The Jacobian for \eqref{eq:Hij2} is computed in Appendix \ref{app:Jacobian},
\begin{equation} \label{eq:Jac}
    \CJ_H 
    = \bigl[\det (-\p^2)\bigr]^2.
\end{equation}
We can eliminate this Jacobian altogether by changing variables from $\eta$ and $\sigma$ to
\begin{align}
    \tilde{\eta} &= \partial^2 \eta, &%
    \tilde{\sigma} &= \partial^2 \sigma.
\end{align} 
The action for $\tilde{\eta}$ is just
\be \label{eq:Seta}
    S_{\tilde{\eta}} = \frac{1}{2} \int dt \, d^2 \mathbf{x} \, \tilde{\eta} \, \qM_{\tilde{\eta}} \tilde{\eta},
\ee 
where
\be 
    \qM_{\tilde{\eta}} = - \partial_{t}^{2} - \qao \partial^4 + 2 \Lambda = \tilde{\CO}_g + 2 \Lambda.
\ee 
Therefore, the contribution of $\tilde{\eta}$ to the partition function is
\be \label{eq:Zeta}
    \mathcal{Z}_{\tilde{\eta}} = \Bigl( \det \qM_{\tilde{\eta}} \Bigr)^{-1/2} = \frac{1}{\sqrt{\det \bigl( \tilde{\mathcal{O}}_g + 2 \Lambda \bigr)}}.
\ee 
Meanwhile, $h$ and $\tilde{\sigma}$ remain coupled via the action
\be \label{eq:Shsigma}
    S_{h\tilde{\sigma}} = \frac{1}{2} \int dt \, d^2 \mathbf{x} 
\begin{pmatrix}
    h & \tilde{\sigma}
\end{pmatrix}
    \qM_{h \tilde{\sigma}}
\begin{pmatrix}
    h \\
    \tilde{\sigma}
\end{pmatrix},
\ee 
where
\be
    \qM_{h\tilde{\sigma}} = \frac{1}{2}
\begin{pmatrix}
    - \bigl( \frac{1}{2} - \lambda \bigr) \partial_{t}^{2} - \bigl [ \gamma + 2 \bigl( \frac{1}{2} - \lambda \bigr)^2 \qac \bigr ] \partial^4 \quad & \bigl [ \gamma - \bigl( \frac{1}{2} - \lambda \bigr) \qac \bigr ] \partial^4 \\[10pt]
    \bigl[ \gamma - \bigl( \frac{1}{2} - \lambda \bigr) \qac \bigr] \partial^4 \quad & - \frac{1}{2} \bigl [ \partial_{t}^{2} + (2 \gamma + \qac ) \partial^4 - 2 \Lambda \bigr ]
\end{pmatrix}.
\ee 
The matrix $\qM_{h\tilde{\sigma}}$ is diagonal only for the choice 
\be
    \qac = \frac{\gamma}{\frac{1}{2} - \lambda}
    \,.
\ee
For general gauge parameters, $\qM_{h\tilde{\sigma}}$ can't be diagonalized locally. 
When $\Lambda = 0$, however, the determinant itself factorizes neatly,
\begin{align} \label{eq:Zhsigma}
    \mathcal{Z}_{h \tilde{\sigma}} \Bigr|_{\Lambda = 0} &= \frac{1}{\sqrt{\bigl ( \det \mathcal{O}_g \bigr ) \bigl ( \det \mathcal{O}_\text{phys} \bigr )}},
\end{align}
where
\be 
    \mathcal{O}_\text{phys} = - \bigl( \tfrac{1}{2} - \lambda \bigr) \partial_{t}^{2} - 2 \gamma \bigl( 1 - \lambda \bigr) \partial^4.
\ee 
The operator $\mathcal{O}_\text{phys}$ is independent of the gauge-fixing parameters $\qao$ and $\qat$.

In summary, the modes corresponding to the various dispersion relations are
\begin{subequations}
\begin{align}
    h, \tilde{\sigma}, n_i\ \text{and ghost} &: \mathcal{O}_g = - \partial_{t}^{2} -2 \bigl( 1 - \lambda \bigr) \qac \partial^4, \\
    \tilde{\eta}, n_i\ \text{and ghost} &: \tilde{\mathcal{O}}_g = - \partial_{t}^{2} - \qao \partial^4, \\
    h, \tilde{\sigma} &: \mathcal{O}_\text{phys} = - \bigl( \tfrac{1}{2} - \lambda \bigr) \partial_{t}^{2} - 2 \gamma \bigl( 1 - \lambda \bigr) \partial^4.
\end{align}
\end{subequations}
For these to have the right sign dispersion relation requires
\be \label{eq:cnda1a}
    \qao > 0, 
        \qquad
    (1-\lambda) \qac > 0. 
\ee
Note that the 
``nice gauge'' of \cite{Blas} is when all three of the dispersions, including the unphysical ones, are actually identical. This condition is satisfied if and only if
\begin{align} \label{eq:nicegauge}
    \qao &= 2 \gamma \frac{1 - \lambda}{\frac{1}{2} - \lambda}, &%
    \qac &= \frac{\gamma}{\frac{1}{2} - \lambda}.
\end{align} 

Finally, the total on-shell partition function is the product of \eqref{eq:Zghflat}, \eqref{eq:Zn}, \eqref{eq:Zeta}, \eqref{eq:Zhsigma} and the extra factor of $( \det \CD_{ij} )^{-1/2}$ from integration over $\Phi_i$. The result simplifies greatly,
\be \label{eq:Zflat}
    \mathcal{Z} \bigr|_{\Lambda = 0} = \frac{1}{\sqrt{\det \mathcal{O}_\text{phys}}}.
\ee 
This represents one physical degree of freedom, with dispersion relation 
\be \label{eq:dispersion_flat}
    \omega^2 = 2 \gamma \, \frac{1 - \lambda}{\frac{1}{2} - \lambda} \, k^4.
\ee 
This dispersion relation is healthy when $\lambda > 1$ or $\lambda < \frac{1}{2}$.
Note that this degree of freedom is a linear combination of $h$ and $\tilde{\sigma}$. Therefore, it will not be captured entirely if one neglects everything except the trace component of $h_{ij}$, as was done in \cite{Benedetti}. 
When $\lambda>1$ (and $\gamma > 0$), the overall sign in front of \eqref{eq:Zflat} is negative and we must Wick rotate the field\footnote{In general, this field is some combination of $h$ and $\tilde{\sigma}$. In the ``nice'' gauge \eqref{eq:nicegauge}, this field is just $h$.} corresponding to $\CO_{\text{phys}}$. 

Once again, this dispersion relation can be derived without regard to a specific gauge-fixing procedure, as in the case of the $U(1)$ gauge theory. Details are given in Appendix \ref{app:direct_eigenvalues}.

\subsection{One-loop effective action with a nonzero cosmological constant}
\label{sec:flat space loop}

Let us calculate the contribution to the determinants of first order in $\Lambda$. 
Our object of study is the effective action $\Gamma(\varphi)$, where $\varphi$ denotes the expectation values of all fields $\Phi$ of the gravitational theory. 
Expanding in $\hbar$,
\be
\Gamma (\varphi) = S (\varphi) + \hbar \Gamma_1 (\varphi) + O (\hbar^{3/2})
\,,
\ee
by standard methods the one-loop quantum effective action takes the form%
\footnote{This is not sufficient to define a gauge invariant effective action \cite{Vilkovisky}. The full treatment of defining a gauge invariant off-shell effective action is beyond the scope of this paper. Instead, we will make use of a field redefinition $\zg_{ij} \rightarrow C \, \zg_{ij}$, which will turn out to be sufficient for our purposes.}
\be
\Gamma_1 (\varphi) = \frac{i}{2}\tr\log S^{(2)}(\varphi) \,,
\ee
where 
\be
    S^{(2)}(\varphi)=\frac{\delta^2 S(\Phi)}{\delta \Phi \delta \Phi} \bigg |_{\Phi = \varphi}
\ee
is the second functional derivative of $S$.
Since this is a gauge theory, we must also include ghost contributions after gauge-fixing, leading to the standard expression
\be
\Gamma_1(\varphi) = \frac{i}{2}\text{tr}\log S^{(2)} - i\,\text{tr}\log \CD_{\rm ghost} \,.
\label{eq:effective action}
\ee
Note that
the only dimensionful parameter present is $\Lambda$ itself, with dimension $[\Lambda]=2$.
As a result, the only contribution $\Lambda$ can have to the logarithmic divergence
(and therefore to the one-loop beta functions) is proportional to $\Lambda$.
To evaluate it, it therefore suffices to evaluate the first derivative of
the partition function $\CZ$ with respect to $\Lambda$.
Separating out the $\Lambda$ dependence of $S^{(2)}$, 
\be
    S^{(2)} = M + \Lambda M^{(\Lambda)} 
    \,,
\ee
the fact that $M$ and $M^{(\Lambda)}$ commute allows us to write
\be
\log\det S^{(2)} = \tr\log S^{(2)} 
= \tr\log M + \Lambda\,\tr \bigl ( M^{-1} M^{(\Lambda)} \bigr ) + O(\Lambda^2) \,.
\ee
$M^{(\Lambda)}$ has contributions only from the $\tilde\eta$ and $(h,\tilde\sigma)$ sectors.
Collecting the corresponding objects, from \eqref{eq:Seta} and \eqref{eq:Shsigma}, we have
\be
    M = 
        \begin{pmatrix}
            M_{\tilde{\eta}} \quad & 0 \\[5pt]
            0 \quad & M_{h \tilde{\sigma}}
        \end{pmatrix},
        \qquad
    M^{(\Lambda)} = 
        \begin{pmatrix}
            M_{\tilde{\eta}}^{(\Lambda)} \quad & 0 \\[5pt]
            0 \quad & M_{h\tilde{\sigma}}^{(\Lambda)}
        \end{pmatrix},
\ee
where
\begin{gather}
M_{\tilde\eta} = -\p_t^2 - \qao\partial^4, 
    \qquad
M^{(\Lambda)}_{\tilde\eta} = 2,
    \qquad
M^{(\Lambda)}_{h\tilde{\sigma}} = \frac{1}{2}
    \begin{pmatrix}
        0 \quad & 0 \\
        0 \quad & 1
    \end{pmatrix}, \\[10pt]
M_{h\tilde{\sigma}} = \frac{1}{2}
\begin{pmatrix}
    - \bigl( \frac{1}{2} - \lambda \bigr) \partial_{t}^{2} - \bigl [ \gamma + 2 \bigl( \frac{1}{2} - \lambda \bigr)^2 \qac \bigr ] \partial^4 \quad & \bigl [ \gamma - \bigl( \frac{1}{2} - \lambda \bigr) \qac \bigr ] \partial^4 \\[10pt]
    \bigl[ \gamma - \bigl( \frac{1}{2} - \lambda \bigr) \qac \bigr] \partial^4 \quad & - \frac{1}{2} \bigl [ \partial_{t}^{2} + (2 \gamma + \qac ) \partial^4 \bigr ]
\end{pmatrix}.
\end{gather}
Evaluating the relevant traces, we obtain the integral form
\be
\text{tr} \bigl [ M^{-1}M^{(\Lambda)} \bigr ] = \int dt \, d^2 \mathbf{x} \sum_{I=1}^3 A_I
\int \frac{d\omega\,d^2\mathbf{k}}{(2\pi)^3}G_I(\omega,\bk),
\ee
with propagators
\be
G_I(\omega,\bk) = \frac{1}{\omega^2 - \alpha^2_I k^4}
\ee
and constants
\be
\alpha_1^2 = \qao,
\qquad
\alpha_2^2 = 2\qac (1-\lambda),
\qquad
\alpha_3^2 = 4\gamma\frac{1-\lambda}{1-2\lambda};
\ee
\be
A_1 = 2,
\qquad
A_2 = \frac{1}{1-\lambda},
\qquad
A_3 = \frac{1-2\lambda}{1-\lambda} \,.
\ee
Here, $I=1$ corresponds to the $\tilde\eta$ contribution, while $I=2,3$ arise from the $(h,\tilde\sigma)$ sector.

Later on in this paper we will use heat kernel methods, which preserve diffeomorphism invariance.
It is difficult to use the heat kernel here, however, because we have not diagonalized $M_{h,\tilde\sigma}$.
(Note, however, that in the diagonal ``nice'' gauge this is not a problem.) 
Although it breaks gauge symmetry and modifies the infrared behavior of the
theory, to extract the coefficient of the logarithmic divergence it suffices 
to use a cutoff regularization.
We integrate over all $\omega$ and introduce a cutoff $k_*$ in $k$.
In addition, the denominators have an implicit $+i\varepsilon$, specifying the appropriate Wick rotatation $\omega = i\omega_E$.
The integrals evaluate to
\be
    \int \frac{d\omega\,d^2 \bk}{(2\pi)^3} \, G_I(\omega,\bk)
    = \frac{1}{4\pi i}\frac{\log k_*}{\alpha_I} + \textrm{(finite)} 
    \,,
\ee
giving the final result
\be
\frac{\p}{\p\Lambda}\log\det S^{(2)}\vert_{\Lambda=0,{\rm div}}
= 
\frac{\log k_*}{4\pi i}\lc
\frac{2}{\sqrt{\qao}}
+ \frac{1}{1-\lambda}\frac{1}{\sqrt{2\qac (1-\lambda)}}
+ \frac{1}{2\sqrt\gamma}\left(\frac{1-2\lambda}{1-\lambda}\right)^{3/2}
\rc \,.
\ee
The contribution of $\Lambda$ to the effective action \req{eq:effective action}
is therefore
\be \label{eq:1divLambda}
\Gamma_{1,{\rm div}} (\zR = 0)
= 
\frac{\log k_*}{8\pi}\lc
\frac{2}{\sqrt{\qao}}
+ \frac{1}{\sqrt{2\qac}}\frac{1}{(1-\lambda)^{3/2}}
+ \frac{1}{2\sqrt\gamma}\left(\frac{1-2\lambda}{1-\lambda}\right)^{3/2}
\rc \, \int dt \, d^2 \mathbf{x} \, \Lambda.
\ee
This is obviously gauge-dependent.
As we will discuss in Section~\ref{sec:renorm}, this gauge dependence should be eliminated by a field strength redefinition for the background metric $\zg_{ij}$,
\be
    \zg_{ij} \rightarrow C \, \zg_{ij}.
\ee
We will utilize this field redefinition in the subsequent section in order to extract the key gauge-independent information.

\section{Time-independent Curved Background}
\label{sec:curved background}

In time-independent backgrounds ($\dot{\zg}_{ij} = 0$), the background values of the extrinsic curvature and the Riemann tensor are $\zK_{ij} = 0$ and $\zR^i_{\phantom{i}jk\ell} = \zR^i_{\phantom{i}jk\ell} (\mathbf{x})$, respectively. 
In two spatial dimensions, the Riemann tensor is determined by the scalar curvature,
\be
	\zR^i_{\phantom{i}jk\ell} (\mathbf{x}) = \frac{1}{2} \zR \bigl [ \delta^i_k \, \zg_{j\ell} (\mathbf{x}) - \delta^i_\ell \, \zg_{jk} (\mathbf{x}) \bigr ].
\ee
By dimensional analysis, the only divergence sensitive to $\zD_i \zR$ that can appear in the effective action is $\zB \hspace{0.5mm} \zR$, which is a total derivative.
Therefore, it suffices to take $\zR$ to be constant.

Consider the action \eqref{eq:action} with the coupling constant $\rho$ set to zero,
\be
\label{eq:actionrhozero}
    S = \frac{1}{\kappa^2} \int dt \, d^2 \mathbf{x} \, N \sqrt{g} \bigl \{ K_{ij} K^{ij} - \lambda K^2 - \gamma R^2 - 2 \Lambda \bigr \}.
\ee
We now expand each term in this action to quadratic order in $\kappa$. 
With
\be
	K_{ij} = \frac{1}{2} \kappa \lr \dot{h}_{ij} - \zD_i n_j - \zD_j n_i \rr + O \! \lr \kappa^2 \rr,
\ee
we have
\begin{subequations}
\begin{align}
	N \sqrt{g} K_{ij} K^{ij} & = \frac{1}{4} \kappa^2 \sqrt{\zg} \lr \dot{h}_{ij} \dot{h}^{ij} - 4 \zD^i n^j \dot{h}_{ij} + 2 \zD^i n^j \zD_i n_j +2 \zD^i n^j \zD_j n_i \rr + O \! \lr \kappa^3 \rr, \\[2pt]
	N \sqrt{g} K^2 & = \frac{1}{4} \kappa^2 \sqrt{\zg} \lr \dot{h}^2 - 4 \dot{h} \zD^i n_i + 4 \zD^i n_i \zD^j n_j \rr + O \! \lr \kappa^3 \rr.
\end{align}
\end{subequations}
Moreover,
\begin{align}
& \quad N \sqrt g\, R^2 \notag \\
& = \sqrt{\zg} \, \zR^2 + \kappa \sqrt{\zg} \, \zR \ls 2\zD^i(\zD^j h_{ij} - \zD_i h) - \tfrac{1}{2}\zR \, h \rs \notag \\
& \quad + \kappa^2 \sqrt{\zg} \, \bigg \{ \!
\bigl( \zD{}^i\zD{}^jh_{ij} - \zD^2h \bigr)^2 
+ \zR^2 \bigl ( \tfrac{3}{4}h^{ij}h_{ij} - \tfrac{1}{8}h^2 \bigr) \notag \\
& \hspace{2.2cm} + \zR \, \Big [ \tfrac{3}{2}\zD{}_kh_{ij}\zD{}^kh^{ij}
- \zD_i h_{jk} \zD{}^k h^{ij} + 2\zD{}_i h \zD{}^jh^i_j
- 2(\zD{}^jh_{ji})^2 \notag \\
& 
\hspace{3.2cm} - \tfrac{1}{2} \lr \zD_i h \rr^2 + 2h^{ij}\zD{}_i\zD{}_j h - 2h^{ij} \bigl ( \zD{}_j\zD{}_k h^k_i 
+ \zD{}_k\zD{}_j h^k_i 
- \zD^2 h_{ij} \bigr ) \Big ] 
\bigg \} \notag \\
& \quad + O(\kappa^3)\,,
\end{align}
and
\be
N \sqrt{g}\, \Lambda = \sqrt{\bar g} \, \Lambda\lc
1 + \tfrac{1}{2}\kappa\,h
+ \tfrac{\kappa^2}{8} \lr h^2 - 2 h^{ij}h_{ij} \rr
\rc 
+ O(\kappa^3).
\ee

The action can be put on-shell by imposing the equation of motion for the background field $\zg_{ij}$. This essentially sets the cosmological constant to be
\be \label{eq:CC}
	\Lambda = \frac{1}{2} \gamma \overline{R}^2.
\ee
Plugging \eqref{eq:CC} back into the action eliminates the tadpole terms linear in $\kappa$.
The on-shell action is
\be
    S = \frac{1}{\kappa^2} \int dt \, d^2 \mathbf{x} \, N \sqrt{g} \lc K_{ij} K^{ij} - \lambda K^2 - \gamma \bigl ( R^2 + \zR^2 \bigr ) \rc.
\ee
Considering such an on-shell action enables us to observe explicit cancellations between ghosts and non-physical modes, reducing the computation of the effective action to that of a single scalar functional determinant.

We offer one caveat: since we do not impose the constraint equation generated by the gauge choice $n=0$, by ``on shell'' we actually mean the background satisfies the $g_{ij}$ equations of motion.
To render the background~\eqref{eq:CC} fully on-shell requires imposing the further condition $\rho=2\gamma\zR$.
As noted in the introduction, however, neither the lapse nor the value of $\rho$ affects the local divergences, and therefore we can ignore both in the computation at hand.

\subsection{BRST quantization}
\label{sec:BRSTquantization}
%
We now turn to the problem of gauge-fixing.
We will apply the BRST formalism.
Instead of classifying the most general gauge-fixing conditions, let us take a more minimalistic approach and construct a gauge-fixing condition such that the cross terms between $n_i$ and $h_{ij}$ cancel in the BRST action. For this purpose, it is sufficient to set the gauge-fixing functional $f_i$ to
\be \label{eq:fic}
    f_i = \dot n_i - D_1 \zD^j h_{ij} - D_2 \zD_i h 
    \,.
\ee
Here $D_1$ and $D_2$ are local operators of dimension one, which we will take to be linear combinations of the diffeomorphism-invariant objects $\zR$ and $\zB \equiv \zD_i \zg^{ij} \zD_j$. 
As we reviewed in Section~\ref{sec:flat space}, equation~\eqref{eq:fic} is not the most general gauge choice consistent with background diffeomorphism invariance: for example, one may also include in $f_i$ terms of the form
\be \label{eq:DzDzDh}
	\zD_i \zD_j \zD_k h^{jk}
	\,.
\ee
In the zero curvature limit, this extra term is the same as the $\qat$ term in \eqref{eq:flatfic}. In the flat case, if one requires that nonphysical modes have a right sign dispersion relation, the conditions derived in \eqref{eq:cnda1a} must be satisfied. For $\lambda > 1$, a nonzero $\qat$ is indispensable for these conditions to hold. On an on-shell background, the one-loop contributions from nonphysical modes cancel exactly in the partition function, and it is not necessary to include \eqref{eq:DzDzDh} in the gauge-fixing condition. When on-shell, we can focus on $\lambda < \frac{1}{2}$ and adopt the simpler gauge-fixing condition in \eqref{eq:fic}.
Evaluating the partition function will result in a gauge invariant expression that is analytically continuable to $\lambda > 1$.

BRST quantization proceeds by introducing a ghost field $c^i$ associated to the generator of infinitesimal diffeomorphisms.
The BRST differential $s$ acts on the physical fields in the same way as the linearized diffeomorphisms in \eqref{eq:gauge}:
\be
    s n_i = \dot{c}_i + O(\kappa),
        \qquad
	s h_{ij} = \zD_i c_j + \zD_j c_i + O(\kappa).
\ee
We also require a cohomologically trivial BRST pair $(b_i,\Phi_i)$, with fermi and bose statistics respectively.
The ghost sector has BRST variations
\be
	s b_i = \Phi_i \,, 
		\qquad 
	s \Phi_i = 0 \,,
	    \qquad
    s c_i = O(\kappa)
    \,.
\ee

Gauge-fixing actions are given by the BRST differential of a gauge-fixing fermion.
We take the gauge-fixing fermion
\be \label{eq:gff}
	\Psi = - \int dt \, d^2 \mathbf{x} \sqrt{\zg} \, b^i \Big \{ \dot{n}_i - D_1 \zD^j h_{ij} - D_2 \zD_i h - \frac{1}{2} \qD \Phi_i \Big \} \,,
\ee
which gives the BRST-exact action 
\begin{align}
    s \Psi & = - \int dt \, d^2 \mathbf{x} \sqrt{\zg} \, \Phi^i \Big\{  \dot{n}_i - D_1 \zD^j h_{ij} - D_2 \zD_i h - \frac{1}{2} \qD \Phi_i \Big\} \notag \\
        & \quad + \int dt \, d^2 \mathbf{x} \sqrt{\zg} \, b^i \lc \ddot{c}_i - D_1 \zD^j \zD_i c_j - D_1 \zB c_i - 2 D_2 \zD_i \zD^j c_j \rc.
\end{align}
This action is associated to a gauge-fixing condition of the form~\req{eq:fic}, except that we have replaced the $\delta$-function type by a gauge of generalized $\CR_\xi$ 
type. 
We have introduced auxiliary fields $\Phi^i$ of dimension $\frac{1}{2}$ and a local operator $\mathcal{D}$ of dimension 1. We choose the following expression for the operator $\mathcal{D}$:
\be
	\qD = - \qao (\zB + \qth \zR).
\label{eq:qD}
\ee
We intentionally keep the real parameters $\qth$ and $\qao$ which depend on the gauge choice. 
Physical results must be independent of their values, giving a check of the final result.

The full BRST-invariant action is
\be
    S_\text{BRST} = S + s \Psi = S + S_\text{g.f.} + S_\text{ghost},
\ee
where
\begin{align} \label{eq:Sgf}
	S_\text{g.f.} = - \int dt \, d^2 \mathbf{x} \sqrt{\zg} \, \Phi^i \Big \{ \dot{n}_i - D_1 \zD^j h_{ij} - D_2 \zD_i h - \frac{1}{2} \qD \Phi_i \Big \},
\end{align}
and
\be \label{eq:ghaction}
	S_\text{ghost} = - \int dt \, d^2 \mathbf{x} \, \Big \{ \dot{b}_i \dot{c}^i - \bigl ( \zD _i D_1 b_j + \zg_{ij} \zD_k D_2 b^k \bigr ) \bigl ( \zD^i c^j + \zD^j c^i \bigr ) \Big \}.
\ee
The BRST partition function is
\be
    \mathcal{Z}_\text{BRST} = \int \mathscr{D} \! \lc n_i, h_{ij}, b_i, c_i, \Phi_i \rc \, e^{i S_\text{BRST}}. 
\ee

Next, let us integrate out the auxiliary field $\Phi_i$ in $S_\text{BRST}$. Since $\Phi_i$ only appears in $S_\text{g.f.}$, we can focus on the following piece in the partition function,
\be \label{eq:ZB}
	\mathcal{Z}_\Phi \equiv \int \mathscr{D} \Phi_i \, e^{i S_\text{g.f.}}.
\ee
Here, $\Phi_i$ is not dynamical and can be eliminated by imposing its equation of motion,
\be
	\Phi_i = \qD^{-1} \lr \dot{n}_i - D_1 \zD^j h_{ij} - D_2 \zD_i h \rr.
\ee
The resulting action after eliminating $\Phi_i$ in $S_\text{g.f.}$ is
\begin{align} \label{eq:Sgf2}
	- \frac{1}{2} \int dt \, d^2 \mathbf{x} \sqrt{\zg} 
	\lr \dot{n}^i - D_1 \zD^j h^i_j - D_2 \zD^i h \rr 
	\qD^{-1} 
	\lr \dot{n}_i - D_1 \zD^k h_{ik} - D_2 \zD_i h \rr. 
\end{align}
From now on, we will take $S_\text{g.f.}$ to denote the expression~\eqref{eq:Sgf2}, even though it is different from the original expression in \eqref{eq:Sgf}. 

Integrating out $\Phi_i$ in the partition function \eqref{eq:ZB} also contributes a functional determinant. 
To evaluate this determinant, we first make the change of variables
\be
	\Phi_i = \zD_i \phi + \ze_{ij} \zD^j \tilde{\phi},
\ee
with $\ze_{ij}=\sqrt{\zg} \, \epsilon_{ij}$
the covariant Levi-Civita symbol for $\bar g$.
The Jacobian is given by \eqref{eq:JPhiapp} in Appendix \ref{app:Jacobian},
\be \label{eq:JB}
    \CJ = \det \! \lr - \zB \rr.
\ee
In terms of $\phi$ and $\tilde{\phi}$, the part of \eqref{eq:Sgf} that is quadratic in $\Phi_i$ can be written as
\be
	S_{\Phi_i}^{(2)} = \frac{1}{2} \int dt \, d^2 \mathbf{x} \sqrt{\zg} \lc \phi \hspace{0.5mm} \zD_i \qD \zD^i \phi + \tilde{\phi} \hspace{0.5mm} \zD_i \qD \zD^i \tilde{\phi} \rc.
\ee
To derive that the cross term between $\phi$ and $\tilde\phi$ is zero, we used the form~\req{eq:qD} of $\qD$ and applied Identity \ref{id:1} in Appendix \ref{app:uf}. 
Therefore,
\be
    \int \mathscr{D} \{\phi, \tilde{\phi}\} \, e^{i S_{\Phi_i}^{(2)}} = \lr \det{}_{\! \Phi_i} \qD \rr^{-1/2},
\ee
where the functional determinant is evaluated to be
\be \label{eq:ZBr}
	\det{}_{\!\Phi_i}\qD = \ls \det \bigl( \zD_i \qD \zD^i \bigr) \rs^{2}
	\,.
\ee
Therefore, the final expression for the $\Phi_i$ contribution \eqref{eq:ZB} is
\be
	\cZ_\Phi = \CJ^{}_\Phi e^{iS_{\rm g.f.}}
	\,,
\ee
where $S_\text{g.f}$ is given by \eqref{eq:Sgf2} and
\be
    \CJ^{}_\Phi = \frac{\det \! \lr - \zB \rr}{\sqrt{\det_{\Phi_i} \! \mathcal{D}}} = \frac{1}{\bigl ( \det |\qao| \bigr ) \det \! \ls - \zB - \lr \qth + \frac{1}{2} \rr \zR \rs}.
\label{eq:JPhi}
\ee
We applied Identity \ref{id:1} for the second equality in \eqref{eq:JPhi}. 

Finally, we determine the operators $D_1$ and $D_2$ in \eqref{eq:gff} by requiring that the cross terms between $n_i$ and $h_{ij}$ cancel in the sum $S + S_\text{g.f.}$, with $S_\text{g.f.}$ set to the expression in \eqref{eq:Sgf2}. The kinetic contribution in the action $S$ comes from
\begin{align}
	S_\mathcal{K} & = \frac{1}{\kappa^2} \int dt \, d^2 \mathbf{x} \sqrt{g} \, \Big \{ K_{ij} K^{ij} - \lambda K^2 \Big \}. \notag 
\end{align}
The part contained in $S_\mathcal{K}$ that is quadratic in terms of the fluctuations is
\begin{align}
    & \frac{1}{4} \int dt \, d^2 \mathbf{x} \sqrt{\zg} \, \Big \{ \dot{h}_{ij} \dot{h}^{ij} - \lambda \dot{h}^2 - 4 \dot{n}^i \lr \zD^j h_{ij} - \lambda \zD_i h \rr \notag \\
		& \hspace{5.25cm} + 2 \ls \zD^i n^j \zD_i n_j + \zD^i n^j \zD_j n_i - 2 \lambda ( \zD^i n_i )^{\!2} \rs \Big \}.
\end{align} 
The cross terms in $S_\mathcal{K}$ are
\be \label{eq:ctkin}
	- \int dt \, d^2 \mathbf{x} \sqrt{\zg} \, \dot{n}^i \lr \zD^j h_{ij} - \lambda \zD_i h \rr.
\ee
The contributions to the cross terms from $S_\text{g.f.}$ are
\be \label{eq:ctgf}
	\int dt \, d^2 \mathbf{x} \sqrt{\zg} \, \dot{n}^i \lr \qD^{-1}D_1 \zD^j h_{ij} + \qD^{-1}D_2 \zD_i h \rr
\ee
These two contributions, \eqref{eq:ctkin} and \eqref{eq:ctgf}, cancel if
\be
	D_1 = - \frac{1}{\lambda} D_2 = \qD.
\ee
Since $\mathcal{D}$ has been defined in \eqref{eq:qD}, this fixes both $D_1$ and $D_2$.

\subsection{The ghost sector}
%
The integration over the ghosts in the partition function can be treated separately. From \eqref{eq:ghaction} we obtain
\be
	S_\text{ghost} = - \int dt \, d^2 \mathbf{x} \lc \dot{b}_i \dot{c}^i - \lr \zD _i D_1 b_j + \zg_{ij} \zD_k D_2 b^k \rr \lr \zD^i c^j + \zD^j c^i \rr \rc.
\ee
We would like to evaluate the partition function
\be
	\mathcal{Z}_\text{ghost} \equiv \int \mathscr{D} \{b_i, c_i\} \, e^{i S_\text{ghost}}.
\ee
Let us reparametrize the ghosts $c_i$ and the anti-ghosts $b_i$ by
\be
	c_i = \zD_i c + \ze_{ij} \zD^j \tilde{c},
		\qquad
	b_i = \zD_i b + \ze_{ij} \zD^j \tilde{b}.
\ee
Similar to \eqref{eq:JB} but for fermions instead of bosons, these changes of variables give rise to the Jacobian
\be
	\CJ_\text{ghost} = \frac{1}{\bigl[ \det \! \lr - \zB \rr \bigr]^2}.
\ee
In terms of the fields $b$, $\tilde{b}$, $c$ and $\tilde{c}$, the ghost action becomes
\begin{align}
	S_\text{ghost} = \int dt \, d^2 \mathbf{x} \sqrt{\zg} \Big \{ b \, \zB \mathcal{O}_g \, c + \tilde{b} \, \zB \tilde{\mathcal{O}}_g \, \tilde{c} \Big \},
\end{align}
where
\begin{align}
	\mathcal{O}_g & = - \partial_t^2 - 2 \qao \! \ls \lr 1 - \lambda \rr \zB + \tfrac{1}{2} \zR \rs \ls \zB + \lr \qth + \tfrac{1}{2} \rr \zR \rs, \\[5pt]
	\tilde{\mathcal{O}}_g & = - \partial_t^2 - \qao \! \lr \zB + \zR \rr \ls \zB + \lr \qth + \tfrac{1}{2} \rr \zR \rs. 
\end{align}
Therefore,
\be
	\mathcal{Z}_\text{ghost} = 
	    \bigl( \det \mathcal{O}_g \bigr) 
	    \bigl( \det \tilde{\mathcal{O}}_g \bigr).
\ee

\subsection{The non-ghost sector}

Now, we would like to come back to examine the non-ghost part in the action $S_\text{BRST}$, namely, the combined contribution from $S + S_\text{g.f.}$. 

It is useful to take the following decomposition of the metric fluctuation $h_{ij}$ such that
\be
	h_{ij} = H_{ij} + \frac{1}{2} \zg_{ij} h,
\ee
where $H_{ij}$ is a traceless 2-tensor, and
\be \label{eq:dech}
	H_{ij} = H^{\perp}_{ij} + \zD_i \eta^{}_j + \zD_j \eta^{}_i + \zD_i \zD_j \sigma - \frac{1}{2} \zg_{ij} \zB \sigma,
\ee
where
\be
	\zg^{ij} H^{\perp}_{ij} = 0, 
		\qquad
	\zD^j H^{\perp}_{ij} = 0,
		\qquad
	\zD^i \eta^{}_i = 0.
\ee
Note that the quantum field $H^{\perp}_{ij}$ is both traceless and divergenceless. In $2+1$ dimensions, $H^{\perp}_{ij}$ encodes only global information about the geometry of the spatial slice (the moduli of the Riemann surface), and carries no local degrees of freedom. 
Therefore, we can drop $H^{\perp}_{ij}$ without affecting the $\beta$-functions. The constraint on $\eta^{}_i$ can be solved by parametrizing $\eta^{}_i$ as
\be
	\eta^{}_i = \ze_{ij} \zD^j \eta.
\ee
The Jacobian from the transformation \eqref{eq:dech} is computed in \eqref{eq:JHij},
\be
    \CJ^{}_H = \det \bigl [ \zB \lr \zB + \zR \rr \bigr ].
\ee

Under the decomposition \eqref{eq:dech}, we have
\be
	S + S_\text{g.f.} = S_n + S_\eta + S_{h \sigma},
\ee
where
\begin{subequations}
\begin{align}
	S_n & = \frac{1}{2} \int dt \,d^2 \mathbf{x} \sqrt{\zg} \, n_i \lc \zg^{ij} \ls - \qao^{\!-1} \! \lr \zB + \qth \zR \rr^{-1} \partial_t^2 - \zB \rs - \zD^j \zD^i + 2 \lambda \zD^i \zD^j \rc n_j, \label{eq:Sna}\\[5pt]
	S_\eta & = \frac{1}{2} \int dt \,d^2 \mathbf{x} \sqrt{\zg} \, \eta \, \zB \lr \zB + \zR \rr \Big \{ \! - \partial_t^2 - \qao \lr \zB + \zR \rr \ls \zB + \lr \qth + \tfrac{1}{2} \rr \zR \rs \Big \} \eta, 
\label{eq:eta action} \\[5pt]
	S_{h\sigma} & = \frac{1}{4} \int dt \,d^2 \mathbf{x} \sqrt{\zg} \, h \lc - \lr \tfrac{1}{2} - \lambda \rr \partial_t^2 - \gamma \lr \zB + \zR \rr^2 - 2 \qao \lr \tfrac{1}{2} - \lambda \rr^2 \zB \bigl [ \zB + \lr \qth + \tfrac{1}{2} \rr \zR \bigr ] \rc h \notag \\
		& \quad + \frac{1}{8} \int dt \,d^2 \mathbf{x} \sqrt{\zg} \, \sigma \, \zB \lr \zB + \zR \rr \Big \{ \! - \partial_t^2 - 2 \gamma \zB \lr \zB + \zR \rr - \qao \lr \zB + \zR \rr \bigl [ \zB + \lr \qth + \tfrac{1}{2} \rr \zR \bigr ] \! \Big \} \sigma \notag \\
		& \quad + \frac{1}{2} \int dt \,d^2 \mathbf{x} \sqrt{\zg} \, \sigma \, \zB \lr \zB + \zR \rr \left \{ \gamma \lr \zB + \zR \rr - \qao \lr \tfrac{1}{2} - \lambda \rr \bigl [ \zB + \lr \qth + \tfrac{1}{2} \rr \zR \bigr ] \right \} h. 
\label{eq:h sigma action}
\end{align}
\end{subequations}
The full one-loop BRST partition function can be written as 
\begin{align}
	\mathcal{Z}_\text{BRST} = \CJ^{}_\Phi \, \CJ^{}_H \, \mathcal{Z}_\text{ghost} \, \mathcal{Z}_n \, \mathcal{Z}_\eta \, \mathcal{Z}_{h \sigma}, 
\end{align}
where
\begin{align}
	\mathcal{Z}_n = \int \mathscr{D} n_i \, e^{i S_n}, 
		\qquad
	\mathcal{Z}_\eta = \int \mathscr{D} \eta \, e^{i S_\eta},
		\qquad
	\mathcal{Z}_{h \sigma} = \int \mathscr{D} \! \lc h, \sigma \rc \, e^{i S_{h \sigma}}.
\end{align}
For $\qao > 0$ (and $\lambda < \tfrac{1}{2}$), we must Wick rotate $n_i$ as well as the time when performing the path integral.

First, let us focus on $\cZ_n$. 
We decompose $n_i$ into scalar components $\nu$ and $\tilde{\nu}$ as follows,
\be
	n_i = \zD_i \! \ls \zB + \lr \qth + \tfrac{1}{2} \rr \zR \rs \nu + \ze_{ij} \zD^j \! \ls \zB + \lr \qth + \tfrac{1}{2} \rr \zR \rs \tilde{\nu} 
	\,.
\ee
We choose this particular decomposition in order to make the action \eqref{eq:Sna} local.
The corresponding Jacobian is
\be
	\CJ_n = \det \lc \lr - \zB \rr \ls \zB + \lr \qth + \tfrac{1}{2} \rr \zR \rs^2 \rc.
\ee
Under this parametrization, we obtain
\begin{align}
	S_n = - \frac{1}{2 \qao} \int dt \, d^2 \mathbf{x} \sqrt{\zg} \, \bigg \{ \nu \, \zB \ls \zB + \lr \qth + \tfrac{1}{2} \rr \zR \rs \mathcal{O}_g \, \nu + \tilde{\nu} \, \zB \ls \zB + \lr \qth + \tfrac{1}{2} \rr \zR \rs \tilde{\mathcal{O}}_g \, \tilde{\nu} \bigg \}.
\end{align}
Collecting these results gives the partition function of $n_i$,
\be
	\mathcal{Z}_n = \frac{\bigl( \det |\qao| \bigr) \det \! \ls - \zB - \lr \qth + \frac{1}{2} \rr \zR \rs}{\sqrt{\bigl ( \det \mathcal{O}_g \bigr ) \bigl ( \det \tilde{\mathcal{O}}_g \bigr )}}.
\ee
Contributions from $\eta$, $\sigma$ and $h$ can be read off of the actions~(\ref{eq:eta action}-\ref{eq:h sigma action}) (in the $h,\sigma$ sector the differential operator is a $2\times 2$ matrix, whose determinant we take directly) and give
\begin{subequations}
\begin{align}
	\mathcal{Z}_\eta & = \frac{1}{\sqrt{\det \! \ls \zB \, ( \zB+\zR) \rs}} \frac{1}{\sqrt{\det \tilde{\mathcal{O}}_{g}}}, \\
	\mathcal{Z}_{h\sigma} & = \frac{1}{\sqrt{\det \! \ls \zB \, ( \zB+\zR) \rs}} \frac{1}{\sqrt{\bigl ( \det \mathcal{O}_g \bigr ) \bigl ( \det \mathcal{O}_\text{phys} \bigr )}},
\end{align}
\end{subequations}
where
\begin{align}
	\mathcal{O}_\text{phys} & = - \lr \tfrac{1}{2} - \lambda \rr \partial_t^2 - 2 \gamma \lr \zB + \zR \rr \ls \lr  1 - \lambda \rr \zB + \tfrac{1}{2} \zR \rs.
\end{align}
%

\subsection{Reduction to physical spectrum}
Let us collect the results that we have derived above. The BRST partition function $\mathcal{Z}_\text{BRST}$ is given by
\begin{align}
	\mathcal{Z}_\text{BRST} = \CJ^{}_{\Phi} \, \CJ^{}_H \, \mathcal{Z}_\text{ghost} \, \mathcal{Z}_n \, \mathcal{Z}_\eta \, \mathcal{Z}_{h \sigma},
\end{align}
where,
\be
	\CJ^{}_\Phi = \frac{1}{\bigl ( \det |\qao| \bigr ) \det \! \ls - \zB - \lr \qth + \frac{1}{2} \rr \zR \rs},
		\qquad
	\CJ^{}_H = \det \bigl[\zB \lr \zB + \zR \rr\bigr],
\ee
and
\begin{subequations}
\begin{align}
	\mathcal{Z}_\text{ghost} & = \bigl( \det \mathcal{O}_g \bigr) \bigl( \det \tilde{\mathcal{O}}_g \bigr ), \\[5pt]
	\mathcal{Z}_n & = \frac{\bigl( \det |\qao| \bigr) \det \! \ls - \zB - \lr \qth + \frac{1}{2} \rr \zR \rs}{\sqrt{\bigl ( \det \mathcal{O}_g \bigr ) \bigl ( \det \tilde{\mathcal{O}}_g \bigr )}}, \\[5pt]
	\mathcal{Z}_\eta & = \frac{1}{\sqrt{\det \bigl[\zB \lr \zB + \zR \rr\bigr]}} \frac{1}{\sqrt{\det \tilde{\mathcal{O}}_g}}, \\[5pt]
	\mathcal{Z}_{h\sigma} & = \frac{1}{\sqrt{\det \! \ls \zB \lr \zB+\zR \rr \rs}} \frac{1}{\sqrt{\bigl ( \det \mathcal{O}_g \bigr ) \bigl ( \det \mathcal{O}_\text{phys} \bigr )}}.
\end{align}
\end{subequations}
The operators $\mathcal{O}_g$, $\mathcal{O}_{\tilde{g}}$ and $\mathcal{O}_\text{phys}$ take the form
\begin{subequations}
\begin{align}
	\mathcal{O}_g & = - \partial_t^2 - 2 \qao \! \ls \lr 1 - \lambda \rr \zB + \tfrac{1}{2} \zR \rs \ls \zB + \lr \qth + \tfrac{1}{2} \rr \zR \rs, \\[5pt]
	\tilde{\mathcal{O}}_g & = - \partial_t^2 - \qao \! \lr \zB + \zR \rr \ls \zB + \lr \qth + \tfrac{1}{2} \rr \zR \rs, \\[5pt]
	\mathcal{O}_\text{phys} & = - \lr \tfrac{1}{2} - \lambda \rr \! \partial_t^2 - 2 \gamma \! \lr \zB + \zR \rr \ls \lr 1 - \lambda \rr \zB + \tfrac{1}{2} \zR \rs.
\end{align}
\end{subequations}
The full BRST partition function reduces to
\be \label{eq:ZBRSTphys}
	\mathcal{Z}_\text{BRST} = \frac{1}{\sqrt{\det \mathcal{O}_\text{phys}}}.
\ee
It is reassuring that the final result is gauge independent and all singular prefactors simply cancel. This partition function counts exactly one physical degree of freedom. 
On the other hand, on an off-shell background there is no reason to expect the result to reduce to a single functional determinant, and the analysis would be more difficult.

While the preceding discussion is formally correct, some care must be taken with analytic continuation to ensure that the path integral converges properly.
Requiring that $\tilde{\CO}_g$ give rise to a sensible dispersion relation gives $\qao > 0$; for $\mathcal{O}_g$, this requires that $\lambda < 1$.
However, both of these operators drop out in the final BRST partition function, and the singular behavior for $\mathcal{O}_g$ (when $\lambda > 1$) can be fixed by modifying the gauge-fixing condition \eqref{eq:fic}. 
Working on-shell gives us the luxury of ignoring this issue: both the operators $\mathcal{O}_g$ and $\tilde{O}_g$ cancel out in the final BRST partition function.

All that remains is the determinant of $\mathcal{O}_\text{phys}$ in \eqref{eq:ZBRSTphys}, whose evaluation requires an appropriate choice of contour.
The coefficient of $\p_t^2$ in $\CO_\text{phys}$ has a healthy sign for $\lambda<1/2$, in which case the standard contour will do.
For $\lambda > 1$ on the other hand, when we perform Wick rotation we must also rotate the field; this is perhaps not surprising, since a similar rotation must be done for the scale factor in general relativity to get a well-defined Euclidean path integral.

In momentum space, we obtain the following dispersion relation for the physical degree of freedom:
\be
	\omega^2 = 2 \gamma \, \frac{1 - \lambda}{\frac{1}{2} - \lambda} \lr k^2 - \zR \rr \lc k^2 - \frac{1}{2 \lr 1 - \lambda \rr} \zR \rc.
\ee	
Note that there are values such that the right-hand side is negative, indicating instability. 
On the sphere ($\zR>0$), at most one unstable mode can arise, namely the zero-momentum mode which is unstable for $\lambda>1$.%
\footnote{In fact, the zero-momentum mode is always projected out when we take into account the lapse constraint. 
We should note, however, that our background only satisfies the lapse constraint for a particular choice of $\rho$.}
More troubling is the case where $\lambda>1$ and $\bar R<0$, since as $\lambda\to 1^+$, the range of momenta with unstable dispersion will grow arbitrarily large.
Nonetheless, provided that the UV scale is much larger than $\bar R/(1-\lambda)$ this will not affect the divergences of the theory, and so for the purposes of computing the beta function we can ignore any instabilities in the low momentum modes.

\subsection{Evaluation of the heat kernel}
It remains to compute the determinant of~\eqref{eq:ZBRSTphys}, which we will do using zeta function regularization.
The real time quantum effective action is
\be
    \Gamma (\varphi) = S (\varphi) + \hbar \Gamma_1 (\varphi) + O (\hbar^2),
\ee
where
\be
    \Gamma_1 (\varphi) = \frac{i}{2} \text{tr} \log \lc S^{(2)} / k_*^4 \rc,
\ee
and
\be 
	S^{(2)} \equiv - \lr \tfrac{1}{2} - \lambda \rr^{-1} \mathcal{O}_\text{phys} = \partial_t^2 + 2 \gamma \frac{1 - \lambda}{\frac{1}{2} - \lambda} \lr \zB + \zR \rr \ls \zB + \frac{\zR}{2 \lr 1 - \lambda \rr} \rs.
\ee
Here, we have introduced a (spatial) momentum scale $k_*$, with $[k_*] = \frac{1}{2}$. 

The zeta function $\zeta(\qs)$ for the operator $S^{(2)}$ is defined in terms of the eigenvalues $\lambda_m$ of $S^{(2)}$ by
\be
	\zeta (\qs) = k_*^{4s} \sum_m \frac{1}{\lambda_m^\qs}
	\,,
\ee
so that
\be
	\log \det S^{(2)} = - \frac{d}{d\qs} \zeta^{} (\qs) \bigg |_{\qs=0} = - \zeta' (0).
\ee
To evaluate divergences, we will use the standard heat kernel representation
\be
	\zeta^{} (\qs) = \frac{k_*^{4\qs}}{\Gamma(\qs)} \int_0^\infty d\qt \, \qt^{\qs-1} \, \text{Tr} \, e^{- \qt \, S^{(2)}}
	\,,
\ee
which gives us the following representation of the one-loop effective action,
\begin{align}
	\Gamma_1 & = \frac{1}{2i} \zeta' (0) \notag \\
		& = \frac{1}{2} \frac{d}{d\qs} \bigg |_{\qs=0} \frac{k_*^{4\qs}}{\Gamma(\qs)} \int dt \, d^2 \mathbf{x} \, \int_0^\infty d\qt \, \qt^{\qs-1} \, \mathcal{I} (\qt;t,\bx),
\end{align}
where 
\be
	\mathcal{I} (\qt; t,\bx) = - i \, \bra{t, \mathbf{x}} \, e^{-\qt \, S^{(2)}} \, \ket{t, \mathbf{x}}.
\ee
Our background is a product geometry $\R\times M_2$, so we decompose $\ket{t,\bx}=\ket{t}\otimes\ket{\bx}$.
Expanding $\ket t$ in Fourier modes allows us to write
\be
    \CI(\tau;t,\bx) = -i\int\frac{d\omega}{2\pi}e^{i\omega t} e^{-\tau\p_t^2}e^{-i\omega t}\, \CI_{A\CV}(\tau;\bx)
    \,.
\ee
Here we have defined $\CI_{\CO}(\tau;\bx) = \bra{\bx}e^{-\tau \,\CO}\ket{\bx}$
for any spatial differential operator $\CO$ and set
\be \label{eq:AVphys}
	A = 2 \gamma \frac{1 - \lambda}{\frac{1}{2} - \lambda},
		\qquad
	\mathcal{V} = \lr \zB + \zR \rr \ls \zB + \frac{\zR}{2 \lr 1 - \lambda \rr} \rs.
\ee
Note that the $\omega$-integral converges after Wick rotation ($\tilde{t} \equiv i t$, $\tilde{\omega} \equiv - i \omega$). Performing the integral over the frequency,
\be
    \int_{-\infty}^\infty \frac{d\omega}{2\pi} \, e^{- i \omega t} e^{- \qt \partial_t^2} e^{i \omega t} 
    = i \int_{-\infty}^\infty \frac{d \tilde{\omega}}{2\pi} e^{- \qt {\tilde{\omega}}^2}
    = \frac{i}{\sqrt{4\pi \qt}},
\ee
we obtain
\be
	\mathcal{I} (\qt; t,\bx) = \frac{1}{\sqrt{4 \pi \qt}} 
	\CI_{A\CV}(\qt;\bx).
\ee
By rescaling $\qt \rightarrow \qt / A$, we obtain
\be
	\Gamma_1 = \frac{1}{2} \int dt \, d^2 \mathbf{x} \, \frac{d}{d\qs} \bigg |_{\qs=0} \frac{k_*^{4\qs}}{A^\qs \Gamma (\qs)} \int_0^\infty d\qt \, \qt^{\qs-1} \mathcal{I} (\qt;t,\bx),
\ee
and
\be
	\mathcal{I} (\qt;t,\bx) = \frac{A^\frac{1}{2}}{\sqrt{4 \pi \qt}}
	\CI_{\CV}(\tau;\bx)
	\,.
\ee

The spatial term $\CI_\CV$ can be evaluated by using the results of \cite{SG}, which computed the divergent contributions due to operators of the form
\be
	\mathcal{V} = \zB^2 + V^{ij} \zD_i \zD_j + T^i \zD_i + X.
\ee
In our case,
\be
	V^{ij} = \zg^{ij} \zR \, \frac{\frac{3}{2} - \lambda}{1 - \lambda}, 
		\qquad
	T^i = 0, 
		\qquad
	X = \frac{\zR^2}{2 \lr 1 - \lambda \rr}.
\ee
Expanding $\CI_\CV$ in powers of $\tau$ defines the Seeley-Gilkey coefficients,
\begin{align}
    \CI_\CV(\tau;\bx) &= 
    \sqrt{\bar g}\sum_{m=0}^\infty 
    a_m(\bx) \tau^{\frac{m-1}{2}}
    \,.
\end{align}

The logarithmic divergence comes from the $m = 2$ term.
The computation of the Seeley-Gilkey coefficient $a_2$ of \cite{SG} yields for $T^i=0$,
\be
	a^{}_2 = \frac{1}{16\sqrt{\pi}} \lc \frac{1}{16} \lr \zg^{ij} V_{ij} \rr^2 + \frac{1}{8} V_{ij} V^{ij} + \frac{1}{6} \lr \zg^{ij} V_{ij} \rr \zR - \frac{1}{3} V_{ij} \zR^{ij} - 2 X \rc = \frac{\gamma^2 \zR^2}{8 \sqrt{\pi} A^2}.  
\ee
The log divergence can be evaluated by introducing a cutoff $\mu^{-4}$ for the $\qt$-integral, which gives
\be
	\frac{d}{ds} \bigg |_{s=0} \frac{k_*^{4s} A^{\frac{1}{2}-s}}{\Gamma(s)} \int_0^{\mu^{-4}} d\qt \, \qt^{\qs - 1} 
	\rightarrow
	\sqrt{A} 
    \, \log \! \lr \frac{k_*^4}{A \mu^4} \rr 
    + \text{(finite)}
    \,.
\ee
Inserting this into the expression for $\Gamma_1$ gives the one-loop logarithmic divergence of the effective action on our background:
\be \label{eq:1divR2}
	\Gamma_{1, {\rm log}} \bigl(\gamma \zR^2 = 2 \Lambda\bigr) = \frac{\sqrt{2\gamma}}{32 \pi} \lr \frac{\frac{1}{2} - \lambda}{1 - \lambda} \rr^\frac{3}{2} \!\log k_* \! \int dt \, d^2 \mathbf{x} \sqrt{\zg} \, \zR^2.
\ee

\subsection{Renormalization for \texorpdfstring{$\gamma$}{gamma} and \texorpdfstring{$\Lambda$}{Lambda}}\label{sec:renorm}

So far, we have evaluated the one-loop effective action over two different background geometries, both of which are described by a time-independent metric:
\begin{itemize}

\item

The Aristotelian spacetime with a nonzero cosmological constant $\Lambda \neq 0$. This background geometry is off-shell, \textit{i.e.,} the background metric does not satisfy the associated background equations of motion. The effective action was evaluated in \eqref{eq:1divLambda}. The covariant expression is
\be \label{eq:Gamma1divR0}
    \Gamma_{1,{\rm log}} \bigl(\zR = 0\bigr) = Y_\Lambda \int dt \, d^2 \mathbf{x} \, \zN \sqrt{\zg} \, 2 \Lambda,
\ee
where
\be \label{eq:YLambda}
    Y_\Lambda \equiv \frac{1}{16 \pi} \lc \frac{2}{\sqrt{\qao}} + \frac{1}{\sqrt{2\qac}} \frac{1}{(1-\lambda)^\frac{3}{2}} + \frac{1}{2 \sqrt{\gamma}} \lr \frac{1 - 2 \lambda}{1 - \lambda} \rr^\frac{3}{2} \rc \log k_* + O \bigl( \kappa^2 \bigr)
\ee
contains gauge dependence.
Although~\eqref{eq:YLambda} was computed using a sharp cutoff, the coefficient of the logarithmic divergence is universal, so we can use this result in studying the logarithmic divergence that arose in zeta function regularization.

\item

A background geometry with a time-independent metric but a nonvanishing Riemann tensor. We study the on-shell action with $\Lambda$ set to be
\be
    \Lambda = \frac{1}{2} \gamma \zR^2.
\ee
The effective action is given in \eqref{eq:1divR2}:
\be \label{eq:GammalogRLambda}
    \Gamma_{1,{\rm log}} \bigl(\gamma\zR^2 = 2 \Lambda\bigr) = Y \int dt \, d^2 \mathbf{x}  \, \zN \sqrt{\zg} \, \gamma \zR^2,
\ee
where
\be
    Y \equiv \frac{1}{32\pi} \sqrt{\frac{2}{\gamma}} \lr \frac{\frac{1}{2} - \lambda}{1 - \lambda} \rr^\frac{3}{2} \log k_* + O \bigl( \kappa^2 \bigr).
\ee
This result is on-shell, and therefore guaranteed to be gauge-independent.

\end{itemize}

\noindent 
Since $Y_\Lambda$ is gauge-dependent, we cannot use $Y_\Lambda$ by itself to extract physically meaningful information. 
Our goal will be to eliminate this gauge dependence and identify a physical quantity that can be extracted from $Y$.

We begin by examining the effective action evaluated on an off-shell time-independent background.
We expand to one-loop order, keeping only the logarithmic divergence:
\be
    \Gamma = S + \Gamma_{1,\text{log}} + \cdots, 
\ee
where
\be
    S = \frac{1}{\kappa^2} \int dt \, d^2 \mathbf{x} \, \overline{N} \sqrt{\zg} \lc \zK_{ij} \zK^{ij} - \lambda \zK^2 - \gamma \zR^2 - 2 \Lambda \rc
    \,.
\ee
Note that $\zK_{ij} = 0$ for a time-independent background. From 
\eqref{eq:Gamma1divR0} and \eqref{eq:GammalogRLambda},
we obtain
\be
    \Gamma_{1, {\rm log}} = \int dt \, d^2 \mathbf{x} \, \zN \sqrt{\zg} \lc \gamma \! \lr Y - Y_\Lambda \rr \zR^2 + 2 Y_\Lambda \Lambda \rc.
\ee
The effective action $\Gamma$ on a time-independent background can be written as
\be \label{eq:naiveeff}
    \Gamma = \frac{1}{{\kappa}^2} \int dt \, d^2 \mathbf{x} \, \zN \sqrt{\zg} \lc - \gamma \bigl [ 1 - \kappa^2 (Y - Y_\Lambda) \bigr ] \zR^2 - 2 \Lambda \bigl ( 1 - \kappa^2 Y_\Lambda \bigr ) \rc + \cdots.
\ee 

As we noted, the na\"ive off-shell effective action \eqref{eq:naiveeff} depends on our choice of gauge parameters. 
In fact, as a function on the space of background metrics, the effective action is gauge-independent, but the parametrization of field space can depend on gauge.
Such dependence can therefore be removed by a field redefinition. 
(For example, see \cite{DeWitt, Vilkovisky}.)
In general, these field redefinitions could include curvature terms. 
In our case, however, for dimensional reasons it suffices to rescale the metric.
Under the rescaling,
\be \label{eq:fred}
    \zg_{ij} \rightarrow C \, \zg_{ij}
    \,,
\ee
we have
\be
    \sqrt{\zg} \rightarrow C \sqrt{\zg},
        \quad
    \zK_{ij} \rightarrow C \zK_{ij},
        \quad
    \zR \rightarrow C^{-1} \zR,
        \quad
    \Lambda \rightarrow \Lambda.
\ee
Under this rescaling, the effective action becomes
\be \label{eq:Ceff}
    \Gamma = \frac{1}{{\kappa}^2} \int dt \, d^2 \mathbf{x} \, \zN \sqrt{\zg} \lc - C^{-1} \bigl [ 1 - \kappa^2 (Y - Y_\Lambda) \bigr ] \gamma \zR^2 - 2 C \bigl ( 1 - \kappa^2 Y_\Lambda \bigr ) \Lambda \rc + \cdots.
\ee 

To extract beta functions requires specifying a normalization condition that fixes the field rescaling. First, let us choose the normalization condition such that the coefficient of the $\zR^2$ term is set to one. This fixes the field rescaling $C$ to be
\be  \label{eq:Cgk}
    C = \frac{\gamma}{\kappa^2} \bigl [ 1 - \kappa^2 (Y - Y_\Lambda )\bigr ],
\ee 
thereby turning the effective action into 
\be \label{eq:GammaOmega}
    \Gamma = \int dt \, d^2 \mathbf{x} \, \zN \sqrt{\zg} \lc - \zR^2 - 2 (1 - \kappa^2 Y) \Omega \rc + \cdots,
\ee
where we have defined
\be
     \Omega \equiv \frac{\gamma \Lambda}{\kappa^4}.
\ee
Indeed, the gauge-dependent contribution $Y_\Lambda$ drops out altogether from this last expression. The factor $(1-\kappa^2 Y)$ can be absorbed into the renormalization of $\Omega$. We are working in bare perturbation theory, so that the physical coupling $\Omega_\text{ph}$ is related to the bare coupling $\Omega$ by $\Omega_\text{ph} = (1 - \kappa^2 Y) \Omega$. 
Then, the anomalous dimension of $\Omega$ is
\be \label{eq:deltaOmega}
    \delta_\Omega \equiv - \frac{d \log \Omega_\text{ph}}{d \log k_*}
    = \frac{1}{16 \pi} \sqrt{\frac{\kappa^4}{2 \gamma}} \lr \frac{\frac{1}{2} - \lambda}{1 - \lambda} \rr^\frac{3}{2} + O (\kappa^4).
\ee

It is interesting to note that the running of $\Omega$ is independent of any field rescaling defined in \eqref{eq:fred}. A simple analysis is helpful for understanding this observation. Throughout the paper, we have taken the scaling dimensions of time and spatial coordinates to be $-1$ and $- \frac{1}{2}$, respectively. In a more fundamental picture, however, we assign two independent dimensions, $T$ to time, and $L$ to length of space. In this latter convention, we have
\begin{align} \label{eq:dim}
    \text{dim} (\kappa^2) = T^{-1} L^{2},
        \qquad
    \text{dim} (\gamma) = T^{-2} L^4,
        \qquad
    \text{dim} (\Lambda) = T^{-2}.
\end{align}
Therefore,
\be
    \text{dim} ( \Omega ) = T^{-2},
\ee
which suggests that $\Omega$ is independent of a rescaling of spatial coordinates. Further note that the rescaling of $\zg_{ij}$ can be absorbed completely into a rescaling of spatial coordinates. Hence $\Omega$ should not change under the field redefinition of the spatial metric. 

As we have seen in \eqref{eq:deltaOmega}, an off-shell time-independent background provides us with only one piece of RG information. There are, however, three couplings, $\kappa$, $\gamma$ and $\Lambda$, in the action evaluated on a time-independent background. Since we have the freedom of choosing a normalization condition to fix the field redefinition, not all these three couplings are independent. By an appropriate choice of the normalization condition, we can at least separate the flow of one coupling constant. Again, we would like to adapt a normalization condition to the spatial curvature term and extract the beta function for the cosmological constant.

Instead of using \eqref{eq:Cgk}, let us first take $C$ to be
\be
    C = \kappa^2 \bigl [ 1 + \kappa^2 C_1 + O (\kappa^4) \bigr ],
\ee
thereby turning the effective action \eqref{eq:Ceff} into
\be
    \Gamma = \int dt \, d^2 \mathbf{x} \, \zN \sqrt{\zg} \lc - \bigl [ 1 - \kappa^2 (Y - Y_\Lambda + C_1) \bigr ] \frac{\gamma}{\kappa^4} \zR^2 - 2 \bigl [ 1 - \kappa^2 (Y_\Lambda - C_1) \bigr ] \Lambda \rc + \cdots.
\ee
Note that $\text{dim} ( \gamma / \kappa^4 ) = 1$ by \eqref{eq:dim}, which motivates us to take a simple choice of the normalization condition by fixing $\gamma / \kappa^4$ to be constant at all scales. Then,
\be
    C_1 = Y_\Lambda - Y,
\ee
and the gauge-independent effective action becomes
\be
    \Gamma = \int dt \, d^2 \mathbf{x} \, \zN \sqrt{\zg} \lc - \frac{\gamma}{\kappa^4} \zR^2 - 2 ( 1 - \kappa^2 Y ) \Lambda \rc + \cdots.
\ee
In bare perturbation theory, we require that the physical couplings $\gamma_\text{ph}$, $\kappa_\text{ph}$ and $\Lambda_\text{ph}$ satisfy
\be
    \frac{\gamma_\text{ph}}{\kappa_\text{ph}^4} = \frac{\gamma}{\kappa^4},   
       \qquad
    \Lambda_\text{ph} = (1 - \kappa^2 Y) \Lambda.
\ee
Therefore, the beta function for $\gamma / \kappa^4$ vanishes,
while the anomalous dimension for the cosmological constant is
\be \label{eq:adim}
    \delta_{\Lambda} \equiv - \frac{d \log \Lambda_\text{ph}}{d \log k_*} = \frac{1}{16 \pi} \sqrt{\frac{\kappa^4}{2 \gamma}} \lr \frac{\frac{1}{2} - \lambda}{1 - \lambda} \rr^{\frac{3}{2}} + O ( \kappa^4 ).
\ee
For $\gamma > 0$ and $\lambda > 1$ or $\lambda < \frac{1}{2}$, $\delta_{\Lambda}$ is real and positive. 
It is interesting to note that when $\lambda = \frac{1}{2}$, which is required for Weyl symmetry, $\delta_{\Lambda}$ vanishes at one-loop order. 
When $\lambda$ approaches $1$, which is required for Lorentz symmetry to be realized, the one-loop expression for $\delta_{\Lambda}$ blows up, reflecting the strong coupling problem of the $\lambda\to 1$ limit~\cite{strong-coupling}. 
Of course, we are still far from determining if the theory is asymptotically free. 
One will have to evaluate the heat kernel for time-dependent background geometries to map out the full RG structure. 

As a final comment, we note that there is no logarithmically divergent contribution to the coupling in front of the term
\be \label{eq:NgR}
    \int dt \, d^2 \mathbf{x} \, \zN \sqrt{\zg} \, \zR.
\ee
This can be seen as follows.
Since the UV properties are controlled by the terms with the most derivatives, we can view $\Lambda$ purely as a coupling constant and expand in a power series of $\Lambda$.
Since $\rho$ does not contribute to the differential operator $\CO_\text{phys}$, $\Lambda$ is the only dimensionful parameter that can arise in the one-loop divergence.
The contribution of lowest dimension, linear in $\Lambda$, has dimension two, and so cannot appear in the coefficient for $\zR$.
Hence, \eqref{eq:NgR} cannot appear at all in the logarithmic divergence at one loop.

\section{Discussion}

This paper dealt with the computation of quantum corrections in the simplest version of critical Ho\v{r}ava gravity, the $z=2$ projectable theory in $2+1$ dimensions.
Working in a gauge with two free parameters, we computed the quantum effective action in two different cases.
The first was flat space with $\Lambda\ne 0$; this is an off-shell background, and we saw that the na\"ive result was gauge dependent.
This gauge dependence is however ephemeral: the effective action in gauge theory can be gauge-dependent, provided the gauge dependence can be eliminated by a field redefinition.

On the other hand, for an on-shell background field an infinitesimal field redefinition leaves the value of the action invariant (since the action is stationary under any variation), and therefore the result (if correct) must be gauge independent.
Working on the time-independent on-shell background $\R\times S^2$ or $\R\times H^2$ with $\gamma\zR^2=2\Lambda$, we find a gauge-independent effective action, as expected.
Using this action, we are able to extract one of the one-loop beta functions.

\medskip
The main result of our paper is therefore equation~\eqref{eq:adim}, which captures the flow of the cosmological constant $\Lambda$ at one loop order in $z=2$ Ho\v{r}ava gravity in $2+1$ dimensions, as defined relative to a metric normalization such that $\gamma / \kappa^4$ is constant at all scales.

\medskip
We focused on the flow of this variable for several reasons, which are all rooted in the fact that our computation is based on the effective action for on-shell, time-independent backgrounds.
Working on-shell has several advantages, notably the automatic gauge invariance of the quantum effective action.
We furthermore saw an explicit reduction of the partition function to only the physical degree of freedom in the one-loop partition function. This simplification can be traced to the on-shell condition.
In this way, the computation of the on-shell effective action could be reduced to the functional determinant of a single scalar operator.

Time independence had the further virtue of allowing us to reduce our computations to known properties of the heat kernels of higher order relativistic differential operators.
And as a background field computation, of course, this can all be done using only the divergences in a single ``vacuum bubble'' diagram, without having to compute vertices explicitly.

\subsection*{Towards the full $\beta$ function} 
One pays a price for
working on time-independent backgrounds, however: 
divergences in the effective action proportional to $K_{ij}$ are invisible.
This means that out of the four couplings%
\footnote{There is a fifth, $\rho$, but as we saw above it receives no logarithmic divergences at one loop.} 
of the model --- $\lambda$, $\kappa$, $\gamma$ and $\Lambda$ --- that played a role here, we can only determine the flow of one.
(Note that not all of these coefficients are physically meaningful. For example, in the text we rescaled $g_{ij}$ to make one coupling take a value of our choosing.)

In order to compute the remaining beta functions, one must relax one of these restrictions.
The full computation can in principle be done entirely on-shell, provided we allow time-dependent backgrounds.
This approach runs into one of two possible difficulties.
The first is that of finding explicit classical backgrounds on which to work.
The simplest backgrounds are cosmological backgrounds of FLRW type,
in which case $\zK_{ij}$ is pure trace.
Imposing the trace constraint reduces the number of beta functions that can be computed by one; to obtain the complete flow of the theory would still require backgrounds on which $\zK_{ij}$ is not pure trace. 

If we accept this limitation, we run into the second complication, that in pure Ho\v{r}ava gravity such backgrounds are de Sitter-like.
As a result they suffer from large contributions to the effective action from temporal boundaries (the boundary area grows at about the same rate as the bulk volume), which makes it difficult to distinguish the boundary and bulk contributions to the effective action.

Even after overcoming these difficulties there remains a potentially troublesome point.
Our methods expressed the determinant in the
$(h,\tilde\sigma)$ sector as a product,
\be
    \det\CO_{h \tilde\sigma} = \det(\CO_g\CO_\text{phys}) = (\det\CO_g)(\det\CO_\text{phys}), 
\ee
after which we cancel against $\CO_g$
coming from the ghost sector.
This requires the product identity $\det(AB)=(\det A)(\det B)$, but this identity runs into difficulties in the infinite-dimensional case.
These can be surmounted straightforwardly when $[A,B]=0$ (as was the case for us), but it is more problematic when $[A,B]\ne 0$, as occurs in the time-dependent case, and leads to ambiguities in the result.
(For one discussion of this issue, see~\cite{heatkernel}.)

These problems point to a general need for more flexible methods to compute loop effects in Ho\v{r}ava gravity.
In the end, it may turn out that the only viable method is to work on perturbative backgrounds, performing explicit expansions of the heat kernel of a matrix differential operator.

\subsection*{Generalization to non-projectable and conformal gravity}
For many purposes, the most interesting class of Ho\v{r}ava gravities are the \emph{non-projectable} theories, which relax the constraint $\nabla_i N=0$ and allow $N=N(t,\bx)$ to depend on space.
For example, in phenomenological applications the non-projectable variant requires much less fine tuning to be consistent with observational constraints~\cite{healthyextension, goodbad}.
From a more conceptual point of view, the ``conformal'' variants -- those invariant under anisotropic Weyl symmetry~\cite{mqc} --- are also of considerable interest.
We here briefly summarize the extension of our methods to these models, and discuss some of the new challenges that arise.

The novelty arising in the non-projectable theory is that once $N$ has local fluctuations, it gives rise to a new constraint.
Because the number of additional constraints equals the number of additional fields (one in both cases) the number of propagating degrees of freedom remains unchanged, but the details of the spectrum and the gravitational interaction are modified.

In the computation of the one-loop effective action, the non-projectability leads to two new features that should be handled carefully.
The first is that $\zN$ cannot be set to 1 by a gauge transformation, and therefore needs to be incorporated appropriately into the gauge-fixing conditions.
The second is that the second-class constraint is non-linear, and so its measure needs to be defined carefully.
The question of whether the right approach is to solve directly for the Dirac bracket, or to use the ghost formalism of~\cite{BatalinFradkin}, or whether there exists a simple prescription giving the correct contributions to the path integral, we leave for future work.

Now for the conformal case.
For certain choices of parameters in the gravitational action, an additional local symmetry arises: anisotropic Weyl invariance.
This is a symmetry under a Weyl scaling
\be
	N \mapsto \Omega^z N
	\qquad
	N_i \mapsto \Omega^2 N_i
	\qquad
	g_{ij} \mapsto \Omega^2 g_{ij}
\ee
where $\Omega=\Omega(t,\bx)$ is an arbitrary function.
In this case, at the classical level the second-class constraint of $N$ is replaced by a first-class constraint, which eliminates the scalar degree of freedom entirely.
The question of whether this symmetry can survive at the quantum level is of considerable interest, particularly in $2+1$ dimensions, where conformal Ho\v{r}ava gravity has no propagating degrees of freedom and therefore provides a useful analog of three dimensional Einstein gravity, with its importance in addressing the conceptual issues of quantum gravity.

In some ways, the conformal case bears similarities to the projectable theory, in that we can gauge fix $\zN=1$ if we like.
On the other hand, to answer questions about the preservation of conformal symmetry, it is important to choose a gauge-fixing condition that is invariant under background Weyl transformations.%
\footnote{This is analogous to the situation in relativistic Weyl gravity in $3+1$ dimensions, see~\cite{FradkinTseytlin}.} 
In particular, if we want to study whether Weyl symmetry is anomalous, we should \emph{not} gauge-fix $\zN=1$, and instead work in a more general background gauge.
This requires us to modify the gauge-fixing conditions.

One important difference in the conformal case is that to preserve background Weyl symmetry, the gauge fixing must respect $z=2$ scaling.
The type of gauge fixing used here and in~\cite{Blas} makes this possible.
It is this consideration that initially led us to the gauge-fixing used in this paper.
We note that background Weyl invariance requires some new features in the gauge-fixing condition, in particular in that $\zN$ and $n$ must be included to construct an appropriate Weyl-invariant object.

Beyond its interest as a toy model, the study of the conformal theory is relevant to the problem of quantum membranes~\cite{mqc}. 
The path integral for relativistic quantum membranes is not renormalizable, putting a theory of fundamental relativistic quantum membranes out of reach.
This is reflected in the Polyakov action formalism in the non-renormalizability of three-dimensional gravity.
With $z=2$ scaling, on the other hand, the Polyakov action becomes power-counting renormalizable.
In this picture, the critical membrane theory would become conformal Ho\v{r}ava gravity coupled to a $z=2$ non-linear sigma model.
The crucial question of whether such critical membrane theories exist, or whether a Weyl anomaly spoils criticality, we leave to future research.

\acknowledgments
The authors would like to thank P.~Ho\v{r}ava for useful discussions at various stages during this project.
C.M.T. would also like to thank A. Coates and S. Mukohyama for fruitful interactions.
The work of T.G. was supported by STFC grant ST/L00044X/1. The work by K.T.G. was supported by ERC Advanced Grant 291092 ``Exploring the
Quantum Universe." 
C.M.T. would like to acknowledge the support of the Thousand Young Talents Program and of Fudan University.
Z.Y. is grateful for hospitality of Fudan University during the preparation of this paper. The work of Z.Y. was supported in part by NSF Grant PHY-1521446 and by Berkeley Center for Theoretical Physics.
Z.Y. would also like to thank the BCTP Brantley-Tuttle fellowship for support while this work was completed.

\appendix

\section{U(1) Gauge Theory Partition Function}
\label{app:U1}
\newcommand\quo{\mathcal{U}}

We compute the partition function \eqref{eq:U1Z} of the $U(1)$ gauge theory in Section \ref{sec:U1} for general $\qD$. The ghost piece reads
\be \label{eq:Zgh}
    \mathcal{Z}_{\text{ghost}} = \int \mathscr{D} \{ b, c \} \, e^{i \int dt \, d^D \mathbf{x} \, b \qO c} = \det \qO,
\ee 
where $\qO$ is the generalized d'Alembertian operator
\be
    \qO = - \partial_{t}^{2} - \partial^4 + v^2 \partial^2.
\ee
We perform the integral over $\Phi$ using the action in \eqref{eq:BRSTexactaction} in order to derive the gauge-fixing action,
\be
    S_{\text{g.f.}} = \int dt \, d^D \mathbf{x} \, 
\begin{pmatrix}
    A_0 & A_i
\end{pmatrix}
    \qM_{\text{g.f.}}
\begin{pmatrix}
    A_0 \\
    A_j
\end{pmatrix},
\ee
the matrix $\qM_{\text{g.f.}}$ is given by
\be \label{eq:Mgf}
    \qM_{\text{g.f.}} = 
\begin{pmatrix}
    - \qD^{-1} \qO + \quo \partial^2 \quad & - \quo \partial_j \partial_t \\[5pt]
    - \quo \partial_t \partial_i \quad & \quo^2 \qD \partial_i \partial_j
\end{pmatrix},
\ee
and the operator $\quo$ is defined as
\be
     \quo \equiv - \qD^{-1} ( \partial^2 - v^2 ).
\ee
The contribution of the gauge fields to the partition function is therefore equal to
\be
    \mathcal{Z}_{A} = \frac{1}{\sqrt{\bigl ( \det \qD \bigr ) \bigl [ \det \bigl ( S^{(2)}_\text{\phantom{g.f.}} +\qM_{\text{g.f.}} \bigr ) \bigr ]}},
\ee
where we recall that the $( \det \qD )^{-1/2}$ piece comes from integrating out the auxiliary field $\Phi$.

The operator $\qM_{\phantom{\text{g.f.}}} \!$ is given in \eqref{eq:U2} and $\qM_{\phantom{\text{g.f.}}} \! + \qM_{\text{g.f.}}$ reads
\be \label{eq:M+Mgf}
    \qM_{\phantom{\text{g.f.}}} \! + \qM_{\text{g.f.}} = 
\begin{pmatrix}
    - \qD^{-1} \qO - (1- \quo) \partial^2 \quad & (1- \quo) \partial_j \partial_t \\[5pt]
    (1- \quo) \partial_t \partial_i \quad & \qO \delta_{ij} - \quo (1- \quo) \qD \partial_i \partial_j
\end{pmatrix}.
\ee 
Here we see explicitly the virtue of the choice $\qD = - \partial^2 + v^2$, or $\quo =1$: 
\be
    \qM_{\phantom{\text{g.f.}}} \! + \qM_{\text{g.f.}} \xrightarrow{\qD = - \partial^2 + v^2} \qO 
\begin{pmatrix}
    - \qD^{-1} \quad & 0 \\[5pt]
    0 \quad & \delta_{ij}
\end{pmatrix},
\ee
whence
\be \label{eq:detM}
    \det \bigl( \qM_{\phantom{\text{g.f.}}} \! + \qM_{\text{g.f.}} \bigr) = \bigl ( \det \qD \bigr )^{-1} \bigl ( \det \qO \bigr )^{D+1},
\ee
and
\be
    \mathcal{Z}_{A} = \bigl ( \det \qO \bigr )^{- \frac{D+1}{2}}.
\ee
Combining this with \eqref{eq:Zgh} gives the total partition function
\be
    \mathcal{Z} = \mathcal{Z}_{A} \mathcal{Z}_{\text{ghost}} = \bigl ( \det \qO \bigr )^{- \frac{D-1}{2}}.    
\ee

To calculate $\det \bigl( \qM_{\phantom{\text{g.f.}}} \! + \qM_{\text{g.f.}} \bigr)$ for general $\qD$, we write $\qM_{\phantom{\text{g.f.}}} \! + \qM_{\text{g.f.}}$ in ADM form,
\be \label{eq:MADM}
    \qM_{\phantom{\text{g.f.}}} + \qM_{\text{g.f.}} = 
\begin{pmatrix}
    -\CN^2 + \CN_i \CN^i \quad & \CN_j \\
    \CN_i \quad & \CG_{ij}
\end{pmatrix}.
\ee
Comparing \eqref{eq:MADM} with \eqref{eq:M+Mgf} immediately gives the ``spatial metric"
\be
    \CG_{ij} = \qO \delta_{ij} - \quo (1- \quo) \qD \partial_i \partial_j,
\ee
the inverse of which is given by
\be
    \CG^{ij} = \tilde{\qD}^{-1} \qO^{-1} \ls \qO \delta^{ij} - \quo (1- \quo) \qD ( \delta^{ij} \partial^2 - \partial^i \partial^j ) \rs,
\ee
where
\be
    \tilde{\qD} = \qO - \quo (1- \quo) \qD \partial^2.
\ee
The ``shift" variables with lower and upper indices are
\begin{subequations} \label{eq:shifts}
\begin{align}
    \CN_i &= (1- \quo) \partial_t \partial_i, \label{eq:N_i} \\
    \CN^i &= \CG^{ij} \CN_i = \tilde{\qD}^{-1} (1- \quo) \partial_t \partial^i. \label{eq:N^i}
\end{align}
\end{subequations}
The ``lapse" function is given by
\be \label{eq:lapse}
    \CN^2 = \qD^{-1} {\tilde{\qD}}^{-1} \qO^2.
\ee
The determinant of $\CG_{ij}$ is
\be \label{eq:detg}
    \det \CG_{ij} = \det \ls 1 - \qO^{-1} \quo (1- \quo) \qD \partial^2 \rs ( \det \qO )^{D} = \bigl( \det \tilde{\qD} \bigr) \lr               \det \qO \rr^{D-1}.
\ee
Finally, we obtain
\be \label{eq:detMADM}
    \det \bigl( \qM + \qM_{\text{g.f.}} \bigr) = \det \CN^2 \, \det \CG_{ij}
    = \bigl ( \det \qD \bigr )^{-1} \bigl ( \det \qO \bigr )^{D+1},
\ee
which agrees with \eqref{eq:detM}, as desired.

\section{Physical Modes without Gauge-Fixing}
\label{app:direct_eigenvalues}

The case of flat space is sufficiently simple that we may actually bypass the gauge-fixing procedure in either the $U(1)$ gauge theory or the Ho\v{r}ava gravity theory and still determine the physical modes and dispersion relations.

In the case of $U(1)$ gauge theory, we start with the action \eqref{eq:U1S}, which has not yet been gauge-fixed. This is written as
\be \label{eq:U1S2}
    S = \frac{1}{2} \int dt \, d^D \mathbf{x} 
\begin{pmatrix} A_0 & A_i \end{pmatrix}
    \qM
\begin{pmatrix}
    A_0 \\
    A_j
\end{pmatrix},
\ee 
where
\be \label{eq:U1M}
    \qM = 
\begin{pmatrix}
    - \partial^2 \quad & \partial_j \partial_t \\[2pt]
    \partial_t \partial_i \quad & ( - \partial_{t}^{2} - \partial^4 + v^2 \partial^2 ) \delta_{ij} + ( \partial^2 - v^2 ) \partial_i \partial_j 
\end{pmatrix}.
\ee 
At this point $A_0$ and $A_i$ have different dimensions. We must rescale one to remedy this. The solution was presented in Section \ref{sec:U1}: redefine $A_0$ by a factor of $\sqrt{\CD}$ where $\CD$ is some spatial differential operator of dimension one, namely some linear combination of $\partial^2$ and $v^2$. Thus, $\qM$ becomes
\be \label{eq:U1M'}
    \qM = 
\begin{pmatrix}
    - \CD \partial^2 \quad & \sqrt{\CD} \, \partial_j \partial_t \\[2pt]
    \partial_t \partial_i \sqrt{\CD} \quad & ( - \partial_{t}^{2} - \partial^4 + v^2 \partial^2 ) \delta_{ij} + ( \partial^2 - v^2 ) \partial_i \partial_j 
\end{pmatrix}.
\ee 
While we will use the inspired choice $\CD = - \partial^2 + v^2$, one could use any other linear combination, including just $v^2$. The subsequent conclusions will not change, which is consistent with the $\CD$-independence shown in Appendix \ref{app:U1}.

Fourier transforming $\qM$ gives
\be \label{eq:U1MFT}
    \qM = 
\begin{pmatrix}
    k^2 (k^2+v^2) \quad & - \omega k_j \sqrt{k^2 + v^2} \\[2pt]
    - \omega k_i \sqrt{k^2 + v^2} \quad & ( \omega^2 - k^4 - v^2 k^2 ) \delta_{ij} + (k^2 + v^2 ) k_i k_j
\end{pmatrix}.
\ee 
For an explicit example, take $D=2$, in which case the above matrix is $3 \times 3$ and can be easily diagonalized. The unnormalized eigenvectors and eigenvalues are

\begin{center}
\begin{tabular}{|c|c|}
    \hline
    Eigenvector & Eigenvalue \\
    \hline\hline
    \vspace{-4mm} & \\
    $\bigl( \omega , k_1 \sqrt{k^2 + v^2}, k_2 \sqrt{k^2 + v^2} \bigr)$ & 0 \\[5pt]
    $\bigl( -k^2 \sqrt{k^2 + v^2} , \omega k_1, \omega k_2 \bigr)$ & $\omega^2 + k^4 + v^2 k^2$ \\[5pt]
    $(0, -k_2, k_1 )$ & $\omega^2 - k^4 - v^2 k^2$ \\
    \hline
\end{tabular}
\end{center}

\noindent The first is a zero mode and is unphysical. The second gives an unphysical dispersion relation and is thus also an unphysical mode. Only the third mode is physical. This mode propagates with the dispersion relation we expect from the gauge-fixing procedure and we can also see from the eigenvector that it is precisely the one transverse mode in the $A_i$'s.

In the Ho\v{r}ava gravity case of Section \ref{sec:flat space},
three of the eigenvalues of the corresponding matrix vanish identically. One of the eigenvalues is given by $\frac{1}{2} \omega^2 + k^4$, which gives an unphysical dispersion relation. The last two eigenvalues are
\be 
    ( 1 - \lambda ) \omega^2 - 2 ( \gamma + \lambda - 1 ) k^4 \pm \sqrt{ \lambda^2 \omega^4 + 4 \lambda^2 \omega^2 k^4 + 4 ( \gamma + 1 - \lambda )^2 k^8}.
\ee 
Among the roots, only one gives a physical dispersion relation, namely \eqref{eq:dispersion_flat},
\be 
    \omega^2 = 2 \gamma \, \frac{1 - \lambda}{\frac{1}{2} - \lambda} \, k^4.
\ee 
%

\section{Useful Formulas} \label{app:uf}

In this appendix, we prove a number of formulas which are useful in expanding out the action of Ho\v{r}ava gravity around curved space. The identities are understood to hold under integration and thus we set all total derivatives to zero. Finally, we take the background to have constant curvature $\overline{R}$.

Recall that we decompose the spatial metric fluctuation as
\be 
    h_{ij} = H_{ij} + \frac{1}{2} h \, \overline{g}_{ij},
\ee 
where $h = \overline{g}^{ij} h_{ij}$ and $\overline{g}^{ij} H_{ij} = 0$. Furthermore, we decompose $H_{ij}$ as
\be \label{eq:Hijtrnsf}
    H_{ij} = H_{ij}^{\perp} + \zD_i \eta_j + \zD_j \eta_i + \Bigl ( \zD_i \zD_j - \frac{1}{2} \overline{g}_{ij} \zB \Bigr ) \sigma,
\ee 
where $\overline{g}^{ij} H_{ij}^{\perp} = 0$, $\overline{\nabla}^j H_{ij}^{\perp} = 0$ and $\zD^i \eta_i = 0$. In two dimensions, we set $H^\perp_{ij}$ to zero and $\eta_i = \ze_{ij} \zD^j \eta$ for some scalar $\eta$.

\begin{Identity} \label{id:1}
    $\overline{\square} \zD_i \Phi = \zD_i \bigl( \overline{\square} + \frac{\zR}{2} \bigr) \Phi$, where $\Phi$ is a scalar. 
\end{Identity}

\begin{Identity} \label{id:2}
    $\zD^j H_{ij} = \ze_{ij} \zD^j ( \overline{\square} + \zR ) \eta + \frac{1}{2} \zD_i ( \overline{\square} + \zR ) \sigma$. 
\end{Identity}
This identity is derived below: 
\begin{align}
    \zD^j H_{ij} &= 
    \ze_{jk} \zD^j \zD_i \zD^k \eta + \ze_{ik} \overline{\square} \zD^k \eta + \lr \zD^j \zD_i \zD_j - \tfrac{1}{2} \zD_i \overline{\square} \rr \sigma \notag \\
    &= \ze_{jk} \bigl( \zD_i \zD^j \zD^k + \zR_{\phantom{o} \ell \phantom{o} i}^{k \phantom{o} j} \zD^{\ell} \bigr) \eta + \ze_{ij} \overline{\square} \zD^j \eta + \lr \zD_i \zD^j \zD_j - \zR_{\phantom{o} j \phantom{o} i}^{k \phantom{o} j} \zD_k - \tfrac{1}{2} \zD_i \overline{\square} \rr \sigma \notag \\
    &= \ze_{jk} \tfrac{\zR}{2} \lr \zg^{kj} \zg_{\ell i} - \delta_{i}^{k} \delta_{\ell}^{j} \rr \zD^{\ell} \eta  + \ze_{ij} \overline{\square} \zD^j \eta + \tfrac{1}{2} \zD_i \overline{\square} \sigma - \tfrac{\zR}{2} ( \zg^{kj} \zg_{ji} - \delta_{i}^{k} \delta_{j}^{j} ) \zD_k \sigma \notag \\
    &= \lr \overline{\square} + \tfrac{\zR}{2} \rr \zD^j \eta + \tfrac{1}{2} \zD_i ( \overline{\square} + \zR ) \sigma \notag \\
    &= \ze_{ij} \zD^j ( \overline{\square} + \zR ) \eta + \tfrac{1}{2} \zD_i ( \overline{\square} + \zR ) \sigma.
\end{align}

\begin{Identity} 
    $\zD^i \zD^j H_{ij} = \frac{1}{2} \overline{\square} ( \overline{\square} + \zR ) \sigma$. This follows immediately from Identity \ref{id:2}.
\end{Identity}

\begin{Identity} \label{id:4}
    $\overline{\nabla}_i \overline{\square}^n \overline{\nabla}^i \Phi = \overline{\square} \bigl( \overline{\square} + \frac{\overline{R}}{2} \bigr)^n \Phi$, where $\Phi$ is a scalar and $n$ is a non-negative integer.
\end{Identity}
In practice, we will only need this identity up to $n=2$. However, it is not much more difficult to prove it in general via induction using Identity \ref{id:1}. 

\begin{Identity} \label{id:5}
    $H_{ij} \zD^i \overline{\square}^n \zD_k H^{jk} = \eta \mathcal{O} \eta + \frac{1}{4} \sigma \mathcal{O} \sigma$, where $\mathcal{O} = \overline{\square} ( \overline{\square} + \zR )^2 \bigl( \overline{\square} + \frac{\zR}{2} \bigr)^n$. This follows directly from Identities \ref{id:2} and \ref{id:4}. 
\end{Identity}

\begin{Identity} \label{id:6}
    For vectors $\Phi_i$ and ${\tilde{\Phi}}_i$,
\begin{subequations}\label{eq:Id6}
\begin{align}
    \Phi_i \zD_j \overline{\square}^{n+1} \zD^j {\tilde{\Phi}}^i &= \Phi_i \bigl( \overline{\square} + \tfrac{\zR}{2} \bigr) \zD_j \overline{\square}^{n} \zD^j {\tilde{\Phi}}^i + \zR \bigl( \Phi_{i} \zD^j \overline{\square}^n \zD^i {\tilde{\Phi}}_{j} - \Phi_i \zD^i \overline{\square}^n \zD^j {\tilde{\Phi}}_j \bigr), \\
    \Phi_i \zD^j \overline{\square}^{n+1} \zD^i {\tilde{\Phi}}_j &= \Phi_i \bigl( \overline{\square} + \tfrac{\zR}{2} \bigr) \zD^j \overline{\square}^{n} \zD^i {\tilde{\Phi}}_j + \zR \bigl( \Phi_{i} \zD_j \overline{\square}^n \zD^j {\tilde{\Phi}}^i - \Phi_i \zD^i \overline{\square}^n \zD^j {\tilde{\Phi}}_j \bigr).
\end{align}
\end{subequations}
From \eqref{eq:Id6} we obtain
\begin{align}
	& \quad \Phi_i \bigl ( \zg^{ij} \zD_k \zB^{n+1} \zD^k + \zD^j \zB^{n+1} \zD^i \bigr ) \tilde{\Phi}_j \notag \\
	& = \Phi_i \bigl ( \zB + \tfrac{3}{2} \zR \bigr )^{n+1} \bigl ( \zg^{ij} \zB + \zD^j \zD^i \bigr ) \tilde{\Phi}_j
		- 2 \zR \sum_{m = 0}^n \Phi_i \bigl ( \zB + \tfrac{3}{2} \zR \bigr )^{n-m} \zD^i \zB^{m} \zD^j \tilde{\Phi}_j.
\end{align}
\end{Identity}

\begin{Identity}  \label{id:7}
    Applying Identity \ref{id:6} on the vectors $\eta_i = \ze_{ij} \zD^j \eta$ and $\zD_i \sigma$ yields
\begin{subequations}
\begin{align}
    \eta_i \zD_j \overline{\square}^n \zD^j \eta^i + \eta_i \zD^j \overline{\square}^n \zD^i \eta_j &= \eta \overline{\square} ( \overline{\square} + \zR ) ( \overline{\square} + 2 \zR )^n \eta, \\
    \sigma \zD_i \zD_j \overline{\square}^n \zD^j \zD^i \sigma - \tfrac{1}{2} \sigma \overline{\square}^{n+2} \sigma &= \tfrac{1}{2} \sigma \overline{\square} ( \overline{\square} + \zR ) ( \overline{\square} + 2 \zR )^n \sigma.
\end{align}
\end{subequations}
\end{Identity}

\begin{Identity} \label{id:8} 
    $H_{ij} \overline{\square}^n H^{ij} = 
    2 \eta \mathcal{O} \eta + \tfrac{1}{2} \sigma \mathcal{O} \sigma$, where $\mathcal{O} = \overline{\square} ( \overline{\square} + \zR ) ( \overline{\square} + 2 \zR )^n$. This follows directly from Identity \ref{id:7}. 
\end{Identity}

\section{Jacobians} \label{app:Jacobian}
Here we review the computation of a partition function under the change of variables $\Phi = F\Psi$, for some linear differential operator $F$ that is often, but not always, local.
The path integral for $\Phi$ is defined with respect to a measure on the space of field configurations, which in the background field method we take to be covariant with respect to background diffeomorphisms.
It is natural to define the measure in terms of a metric $G_\Phi$ on this space.
For example, if $\Phi_i$ has a spatial index, the inner product of two infinitesimal variations $\delta\Phi^{(1)}$ and $\delta\Phi^{(2)}$ takes the form
\be
    G_{\Phi}(\delta\Phi^{(1)},\delta\Phi^{(2)})
    = \corr{\delta\Phi^{(1)},\delta\Phi^{(2)}}_\Phi
    = \int dt \, d^2 \mathbf{x} \sqrt{\zg} \, \zg^{ij} \delta\Phi^{(1)}_i \delta\Phi^{(2)}_j \,.
\ee
The path integral can schematically be written%
\footnote{More correctly, $G_\Phi$ should be taken as the metric induced from the canonical path integral by integrating out canonical momenta. This gives $G_\Phi$ in terms of the path integral kinetic term.}
\be
    \int d\Phi \sqrt{\det G_{\Phi}} \, e^{iS_\Phi} 
    \,.
\ee
The metric is not covariant under a change of variables; instead, the measure transforms as
\be
d\Phi\sqrt{\det G_{\Phi}} = d\Psi \sqrt{\det G_{\Psi}} \sqrt{\det \CO_F},
\ee
where
\be 
    \CO_F = G_{\Psi}^{{}^{-1}} F^{{}^\intercal} G^{\phantom{-1}}_{\Phi} \!\! F.
\ee 
%
The operator $\CO_F$ is computed by setting
\be
    \corr{\delta\Phi,\delta\Phi}_{\Phi}
    = \corr{\delta\Psi,\CO_F\delta\Psi}_{\Psi} \,.
\ee
The Jacobian is then expressed as $\CJ_F = \sqrt{\det\CO_F}$.

As an example, let us consider the Jacobian for the transformation
\be
    \Phi_i = \zD_i\phi + \ze_{ij}\zD^j\tilde\phi 
    \,.
\ee
The natural metric for $\Phi_i$ is the one given above, while that for $\phi$ and $\tilde\phi$ is
\begin{subequations}
\begin{align}
    \corr{\delta \phi^{(1)}, \delta \phi^{(2)}}_{\phi} 
    & = \! \int \! dt\,d^2\mathbf{x}\sqrt{\zg} \, \delta \phi^{(1)} \delta \phi^{(2)}, \\
    \corr{\delta \tilde{\phi}^{(1)}, \delta \tilde{\phi}^{(2)}}_{\tilde{\phi}} 
    & = \! \int \! dt\,d^2\mathbf{x} \sqrt{\zg} \, \delta \tilde{\phi}^{(1)} \delta \tilde{\phi}^{(2)}.
\end{align}
\end{subequations}
In the text we are primarily interested in the case where $\zg$ is time-independent and whose spatial slice is a symmetric space.
Then
\be
    \corr{\delta \Phi, \delta \Phi}^{}_{\Phi} 
    = \corr{\delta \phi,-\zB \delta \phi}^{}_\phi + \corr{\delta \tilde\phi,-\zB \delta \tilde\phi}_{\tilde\phi}
    \,.
\ee
Therefore,
\be
    \CO_F = - \zB \begin{pmatrix} 
                \,1 \,\,\, & 0\, \\[1pt] 
                \,0 \,\,\, & 1\, 
            \end{pmatrix},
\ee
and
\be \label{eq:JPhiapp}
    \CJ = \sqrt{\det\CO_F} = \det(-\zB)
    \,.
\ee

For another example, let us conside the Jacobian for the transformation defined in \eqref{eq:Hijtrnsf},
\be 
    H_{ij} = \Bigl ( \ze_{jk} \zD_i \zD^k + \ze_{ik} \zD_j \zD^k \Bigr ) \eta + \Bigl ( \zD_i \zD_j - \frac{1}{2} \overline{g}_{ij} \zB \Bigr ) \sigma 
    \,.
\ee
Note that $H_{ij}$ is traceless.
A natural metric on the space of traceless tensors is
\be
    \corr{\delta H^{(1)}, \delta H^{(2)}}_H = \int dt \, d^2 \mathbf{x} \sqrt{\zg} \, \delta H^{(1)}_{ij} \zg^{ik} \zg^{j\ell} \delta H^{(2)}_{k\ell}.
\ee
For the scalar modes $\eta$ and $\sigma$ we define
\begin{subequations}
\begin{align}
    \corr{\delta \eta^{(1)}, \delta \eta^{(2)}}_\eta 
    & = \! \int \! dt\,d^2\mathbf{x}\sqrt{\zg} \, \delta \eta^{(1)} \delta \eta^{(2)}, \\
    \corr{\delta \sigma^{(1)}, \delta \sigma^{(2)}}_\sigma 
    & = \! \int \! dt\,d^2\mathbf{x} \sqrt{\zg} \, \delta \sigma^{(1)} \delta \sigma^{(2)}.
\end{align}
\end{subequations}
Then, applying Identity \ref{id:8}, we obtain
\be
	\corr{\delta H, \delta H}^{}_{H} = \corr{\delta \eta, 2 \zB \lr \zB + \zR \rr \delta \eta}^{}_{\eta} + \corr{\delta \sigma, \tfrac{1}{2} \zB \lr \zB + \zR \rr \delta \sigma}^{}_{\sigma}.
\ee
(In general there is an $\eta$-$\sigma$ cross-term involving $\zD_i\zR$, but this vanishes on the backgrounds used in this paper.)
Therefore, the associated Jacobian is
\be \label{eq:JHij}
	\mathcal{J}^{}_H = \det \! \Big [ \zB \lr \zB + \zR \rr \! \Big ].
\ee

\bibliographystyle{JHEP}
\bibliography{pHL}

\end{document}